\documentclass[aip,reprint]{revtex4-1}
\usepackage{lipsum} 
\usepackage{graphicx}
\usepackage{siunitx}
\usepackage[dvipsnames]{xcolor}
\usepackage{amssymb,mathtools}

\renewcommand{\refeq}[1]{Eq.~\ref{#1}}
\newcommand{\ls}[1]{_\mathrm{#1}}
\newcommand{\vect}[1]{{\mathbf{#1}}} % math vector (bold)

\usepackage[dvipsnames]{xcolor}
\usepackage{ulem}

\draft % marks overfull lines with a black rule on the right

\begin{document}

%\preprint{}

\title{Aging or DEAD: origin of the non-monotonic response to weak self-propulsion in active glasses}
%\title{Aging or DEAD: origin of the non-monotonic behavior in active glasses}
%\title{Non-monotonic behavior in active glasses} %Title of paper



\author{Natsuda Klongvessa}
%\thanks{}
\altaffiliation{Now in: Faculty of Medicine, Bangkokthonburi University, 10170 Bangkok, Thailand}
\affiliation{School of Physics, Institute of Science, Center of Excellence in Advanced Functional Materials, Suranaree University of Technology, 3000 Nakhon Ratchasima, Thailand}
\affiliation{Universit\'e de Lyon, Universit\'e Claude Bernard Lyon 1, CNRS, Institut Lumi\`ere Mati\`ere, F-69622, Villeurbanne, France}

\author{Christophe Ybert}
\author{C\'ecile Cottin-Bizonne}
\affiliation{Universit\'e de Lyon, Universit\'e Claude Bernard Lyon 1, CNRS, Institut Lumi\`ere Mati\`ere, F-69622, Villeurbanne, France}
\author{Takeshi Kawasaki}
\affiliation{Department of Physics, Nagoya University, 464-8602 Nagoya, Japan}

\author{Mathieu Leocmach}
\affiliation{Universit\'e de Lyon, Universit\'e Claude Bernard Lyon 1, CNRS, Institut Lumi\`ere Mati\`ere, F-69622, Villeurbanne, France}

\date{\today}

\begin{abstract}
Among amorphous states, glass is defined by relaxation times longer than the observation time. 
This nonergodic nature makes the understanding of glassy systems an involved topic, with complex aging effects or responses to further out-of-equilibrium external drivings.
In this respect active glasses made of self-propelled particles have recently emerged as  stimulating systems which broadens and challenges our current understanding of glasses by considering novel internal out-of-equilibrium degrees of freedom.
%
%This nonergodic nature makes the understanding of glass a challenge.
%\tk{Glass is defined as a state in which the relaxation time is longer than the observation time while taking on a random structure.}
%
%\textcolor{green}{One possible way to test glass theories is to set the system even more out-of-equilibrium, for instance by using active glass made of self-propelled particles.}
%\textcolor{purple}{One possible way to test the glass theories is to set the system further out-of-equilibrium, for example by using an active glass composed of self-propelled particles.} 
%
In previous experimental studies we have shown that in the ergodicity broken phase, the dynamics of dense passive particles first slows down as particles are made slightly active, before speeding up at larger activity.
%In previous} studies, we have shown experimentally that beyond ergodicity breaking, the relaxation is slower in a slightly active colloidal system than} in its passive counterpart, whereas more activity enhances relaxation.
Here, we show that this nonmonotonic behavior also emerges in simulations of soft active Brownian particles and explore its cause.
We refute that the Deadlock by Emergence of Active Directionality (DEAD) model we proposed earlier describes our data.
However, we demonstrate that the nonmonotonic response is due to activity enhanced aging, and thus confirm the link with ergodicity breaking.
Beyond self-propelled systems, our results suggest that aging in active glasses is not fully understood.
%Beyond self-propelled systems, our results suggests aging mechanisms specific to glass made of soft particles.}
\end{abstract}


\maketitle %\maketitle must follow title, authors, abstract and \pacs


\section{Introduction}
%\label{}

Half a century ago, \textcite{Goldstein1969} proposed that the relaxation of glassy systems is dominated by free energy barriers high compared to thermal energy. At low enough temperatures or high enough densities some barriers are so high that the system cannot explore its whole energy landscape and becomes nonergodic~\cite{Debenedetti2001,Berthier2011_review}, with the available phase-space region reducing in a non trivial way with temperature or density~\cite{Charbonneau2014,Charbonneau2017_review}.
Nevertheless partial relaxation still occurs in the nonergodic glass state and typically slows down with the `age' of the system, i.e. the time after its preparation in the glass state~\cite{struik1977physical,berthier_theo_2011}. More generally, the system behavior depends on its whole history and preparation protocol.
Therefore probing the glass state is an out-of-equilibrium issue, and has been performed by studying the response to external driving, for instance temperature cycles~\cite{vincent1997SlowDynamicsAging, scalliet2019RejuvenationMemoryEffects} or global mechanical driving~\cite{Viasnoff.2002, Ikeda2013a}.
More recently has emerged a novel class of systems whereby the additional nonequilibrium condition are brought by the self-propulsive properties of each constitutive particles.
%}
The physics of such active glasses~\cite{janssen2017AgingRejuvenationActive, Janssen2019_ActiveGlass} is expected to be relevant to describe biological tissues~\cite{Garcia2015_cell_activejamming,Bi2016PRX}.
Starting from a zoological curiosity such as flocks of birds and schools of fish, active matter has received more attention and is established as a novel area to study new physics in a scope of nonequilibrium systems~\cite{Ramaswamy2010,Vicsek_review2012,Marchetti2013,Bechinger2016}.
In rather dilute conditions, several studies have shown that the state of many active systems can be described by an effective temperature~\cite{Howse_PRL_2007,Tailleur2009,Palacci2010,Ginot2015}.
In previous experimental studies~\cite{Klongvessa2019_PRL,Klongvessa2019_PRE}, some of us have shown that this mapping holds up close to ergodicity breaking but not beyond.
Our main observation is that the relaxation time displays nonmonotonic response with the self-propulsion force.
Starting from the nonergodic passive system, the relaxation time increases when the system becomes weakly self-propelled and then decreases when the activity level is high enough to restore ergodicity.

Other nonmonotonic behaviors have been reported from theoretical~\cite{liluashviliModecouplingTheoryActive2017} and numerical~\cite{debets2021cage} studies of ergodic glassy systems. Indeed, close to the glass transition the dynamics does not depend on a single active parameter but on at least two parameters characterizing the activity, e.g. the persistence time and the Peclet number. For instance, increasing the persistence time can shift the glass transition towards higher or lower densities depending on the effective temperature~\cite{Berthier2017}.
Ultimately, the direction of the shift of glass transition depends on the microscopic details of the activity~\cite{Nandi2018}.

Beyond the glass transition, some recent studies reveal that the phenomenology of passive glass is qualitatively altered by activity: plasticity and turbulence~\cite{mandal_extreme_2020}, aging behavior~\cite{Mandal_Sollich_PRL2020_multi_aging,janzen2021agingthermalglass}, and dynamics heterogeneity~\cite{paul_DH_activeglass_2021}.
However, most of the above effects occur at relatively strong self-propulsion force, and more scarce are the studies that focus on the perturbation of the thermal motion by a weak propulsion force; precisely the regime where we observed experimentally a nonmonotonic response to activity~\cite{Klongvessa2019_PRL,Klongvessa2019_PRE}.

In Ref.~\cite{Klongvessa2019_PRL}, we proposed a model based on a Deadlock due to the Emergence of Active Directionality (DEAD). Briefly, this model is based on a \textcite{Goldstein1969}-like picture with an (effective) single particle trapped in a cage of high free energy barriers.
We consider that a small amount of self-propulsion does not affect the height of the energy barriers.
However, the nature of the motion within the cage affects the frequency at which the particle could attempt to hop out of its cage. 
A Brownian particle randomly explores its cage and can often test weak points of the cage (low energy barriers). 
A particle with persistent directed motion will be able to test a single direction until it reorients. 
At the scale of the cage, persistent motion can be less efficient than Brownian motion, thus the attempt frequency lower and the structural rearrangement at low activities slower than in the equivalent passive system.

%issues with DEAD
Several experimental issues prevented us to test further the DEAD model. 
(i) The system density was inhomogeneous. 
(ii) We had no access to the orientation of the self-propulsion force, which prevents testing the cage exploration scenario.
(iii) We had a poor control on the system age and history, which prevents ensemble averaging and a systematic study on aging behavior.

Furthermore, in 2D systems like in our experiments, long wavelength sound modes~\cite{shiba2016UnveilingDimensionalityDependence}, also called Mermin–Wagner fluctuations~\cite{illing2017MerminWagnerFluctuations}, are expected and may blur our interpretation in term of a local cage. Indeed, in Ref.~\cite{Klongvessa2019_PRE} we found that most of the observed relaxation was not due to structural relaxation but to motion at a scale of a few particles that we interpreted as collective motion but may have been long wavelength sound modes.

In this work, we use numerical simulations to (i) reproduce the nonmonotonic behavior in dense, nonergodic assembly of soft 2D active Brownian particles while controlling for long wavelength sound modes; (ii) overcome experimental limitations, further characterize this phenomenon; (iii) test the DEAD model and alternative explanations.
We provide the simulation details in Section~\ref{sec:method}, confirm the nonmonotonic behavior in Section~\ref{sec:results}, but refute the cage exploration postulated by the DEAD model in Section~\ref{sec:cage_explore}.
In Section~\ref{sec:aging}, we rather link the nonmonotonic behavior to aging, which is a hallmark of glassy dynamics. 
Finally, we discuss our results and conclude in Section~\ref{sec:discussion}.

\section{Simulation details}
\label{sec:method}
\subsection{Equations of motion}

% equations of motion
To simulate our active system, we use 2D Active Brownian Particle (ABP) model~\cite{Fily2012,Volpe2014}, which consists in the overdamped Langevin equation for Brownian particles with an additional active force. 
The equations of motion of each particle $i$ of diameter $\sigma_i$ are given by 
\begin{eqnarray}
\zeta\dot{\mathbf{r}}_i &&=  \boldsymbol{\xi}_i+ \mathbf{F}^{\mathrm{I}}_i +  f\hat{\mathbf{u}}_i,
\label{eq:ABP_dr}
\\ 
\dot{\mathbf{\theta}}_i &&= \eta_i,
\label{eq:ABP_dtheta}
\end{eqnarray}
where $\zeta$ is a drag coefficient.
$\mathbf{r}_i = (x_i,y_i)$ and $\theta_i$ are the particle position and orientation, respectively. 
The term $\boldsymbol{\xi}_i$ stands for the thermal noise of zero-mean and it obeys $\langle \boldsymbol{\xi}_i(t) \boldsymbol{\xi}_j(t^\prime)\rangle = 2k\ls{B}T\zeta\,\delta_{ij}\delta(t-t^\prime)\mathbf{1}$,
where $\delta_{ij}$ and $\delta(t-t^\prime)$ are the Kronecker and Dirac delta function, respectively, $k\ls{B}$ is Boltzmann's constant and $\mathbf{1}$ is the identity matrix.
The interaction term between particles is $\mathbf{F}^{\mathrm{I}}_i(t) = -\Sigma_j\frac{\partial U(r_{ij})}{\partial\mathbf{r}_i}$, where $U(r_{ij})$ is the interaction potential between particle $i$ and $j$ separated by the distance $r_{ij}$. In this work, we use Harmonic sphere potential~\cite{OHern2002, Berthier2009},
$U(r_{ij}) = \epsilon(1-r_{ij}/\sigma_{ij})^2/2$,
with the cutoff range $\sigma_{ij} = (\sigma_i + \sigma_j)/2$. 
The active force is represented by a self-propulsion term $f\hat{\mathbf{u}}_i(t)$, where $f$ is a self-propulsion force of a constant magnitude. $\hat{\mathbf{u}}_i = (\cos{\theta_i},\sin{\theta}_i)$ is a unit vector subjected to rotational diffusion equation (\refeq{eq:ABP_dtheta}). $\eta_i$ is a zero-mean thermal noise that follows $\langle \eta_i(t)\eta_j(t^\prime)\rangle = 2D\ls{R}\delta_{ij}\delta(t-t^\prime) $, and $D\ls{R}$ is a rotational diffusion coefficient.

By this definition, our particles always have both translational and rotational diffusion originating from Brownian thermal noise. The orientation of the particles $\theta_i$ is the self-propulsion direction in the case of active particles, $f>0$. For passive particles, $f = 0$, despite their orientation also changing diffusively, there is no influence from this change on the displacement of the particles.

We simulate $N = 1000$ particles with periodic boundary condition.
To avoid crystallization, particle diameter has 10\% polydispersity with Gaussian distribution centered at $\sigma_0 = 1$. We detect medium-range crystalline order typical of polydisperse glassy systems~\cite{kawasaki2014StructuralEvolutionAging}, but neither long-range crystalline order nor well defined grains and boundaries.
The units of length, energy and time in our simulation are $\sigma_0$, $\epsilon$ and $\sigma_0^2\zeta/\epsilon$, respectively.

% model parameters
\subsection{Model parameters}

% Relation between D_R and D_T
Translational diffusion and self-propulsion are the two mechanisms that govern motion for an active Brownian particle.
They are characterised by translational ($D\ls{T} = k\ls{B}T/\zeta$) and rotational ($D\ls{R}$) diffusion coefficient, respectively.
$D\ls{R}$ sets the rotational time $\tau\ls{R} = D\ls{R}^{-1}$ that marks for an individual particle the crossover between ballistic motion at short times and effective diffusion at long time scales~\cite{Howse_PRL_2007,Volpe2014}.
In experimental realisation of ABP, $D\ls{R}$ is linked to $D\ls{T}$ through a geometric coefficient. For instance for spherical particles of radius $\sigma_0$ rotating in 3D, we have $D\ls{R}\sigma_0^2/D\ls{T} = 3$.
However, the presence of a solid wall can shift this coefficient toward larger values~\cite{Gomez-Solano.2017}.
Although some numerical studies keep $D\ls{R}\sigma_0^2/D\ls{T} = 3$~\cite{DigregorioPRL2018_fullphasediagramABP,Volpe2014}, most treat $D\ls{R}$ as an independent parameter~\cite{mandal_extreme_2020,Mandal_Sollich_PRL2020_multi_aging}.
In the present study, we fix the temperature at $k\ls{B}T = 0.004$ and set the rotational diffusion coefficient such that $D\ls{R}\sigma_0^2/D\ls{T} = 10$, corresponding to a rotational time $\tau\ls{R} = D\ls{R}^{-1} = 25$.
We will discuss further the significance of the value of $D\ls{R}$ in Section.~\ref{subsec:length_and_time_scale}.
%\cyb{Note in addition that experimentally spherical active colloids confined nearby a solid wall indeed experience a shift of the $D\ls{R}/D\ls{T}$ towards higher values due to different impact on both modes of surface-induced motility hindrance \cite{Gomez-Solano.2017}} 

We define the P\'eclet number $\mathrm{Pe} = f\sigma_0/k\ls{B}T$ to describe the activity level of the system.
In this work, we will investigate the system dynamics at various activity levels $\mathrm{Pe}$ and densities $\phi = \pi\sum_i^N \sigma_i^2/4L^2$, where $L$ is the length of the simulation domain.

\subsection{Simulation procedure}
\label{subsec:Sim_procedure}
%Preparation stage
We solve the equation of motion, Eq.~\ref{eq:ABP_dr}-\ref{eq:ABP_dtheta}, using Heun's method with a step size $\delta t = \num{3e-4}$.
Initially, particles are randomly put in space with no self-propulsion force, $f=0$. 
%\del{Then, we let the system evolve for a short period of time to separate overlapped particles. We monitor the potential energy in the system to verify that it has reached a stationary state, which means that particles are not overlapping anymore.
%For all densities, we always perform this initialization stage for $t\ls{ini} = 20$, at which the system has reached a stationary state for some time.
%}
The system quickly relaxes over a short period of time during which originally overlapping particles separate. This rapid evolution is then followed by a much slower one corresponding for instance to aging effects in the glassy state (see further details in Section~\ref{sec:aging}). The cross-over between the initialization stage and the later evolution can be monitored through the system potential energy.
For all densities this initialization stage is set to a fixed duration $t\ls{ini} = 20$ (except in Fig~\ref{fig:newrun}b and d, as will be explained in Sec.~\ref{sec:protocol_dependence}).
After that, we turn on the self-propulsion, $f = \mathrm{Pe}\times k\ls{B}T/\sigma_0$, which defines the origin of waiting times $t\ls{w} = 0$.

Since this preparation protocol involves infinitely fast annealing from random configuration, we expect strong aging effects~\cite{OHern2002, Berthier2009} that will be addressed in Section~\ref{sec:aging}.
The production run starts at the waiting time $t\ls{w} = \num{1.5e4}$ (unless stated otherwise) over the time interval $\Delta t = \num{3e3}$.
If the system has completely relaxed during this time interval (see results in Section~\ref{subsec:results_supercooled}), the simulation results are a time-average over six successive production runs.
Otherwise, they are an ensemble-average over 50 independent runs at the same waiting time.

\section{Active glassy behavior}
\label{sec:results}
\subsection{Dynamics before glass transition}
\label{subsec:results_supercooled}

\begin{figure}
\includegraphics{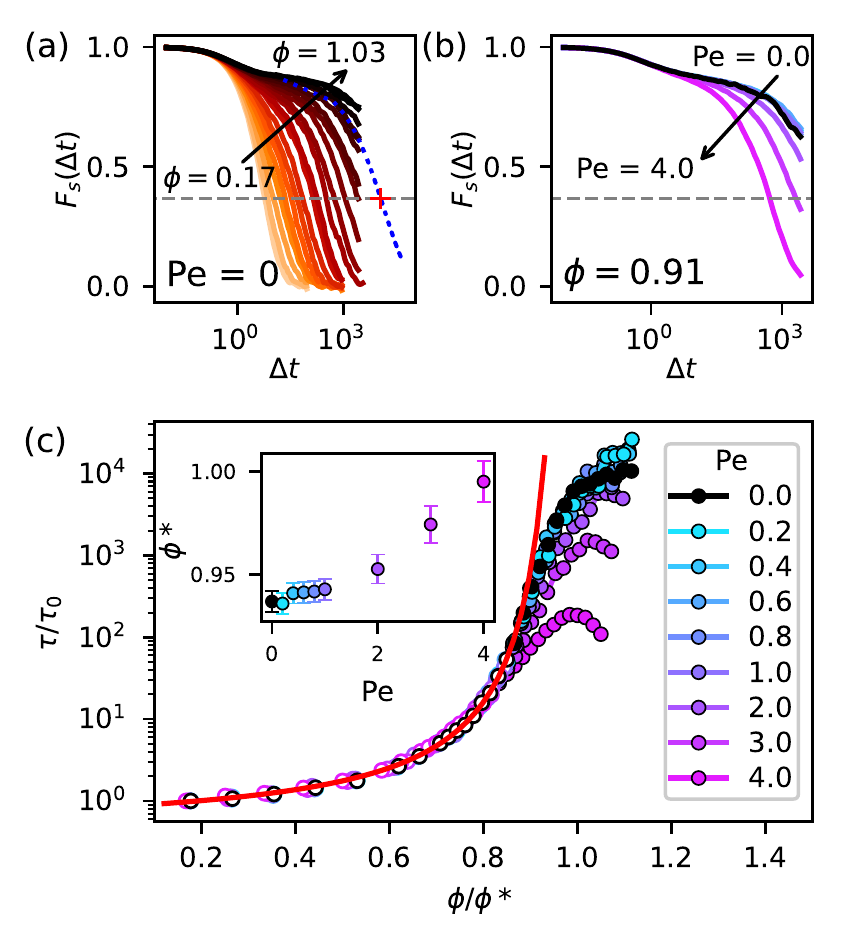}
\caption{
(a,b) Self-part of the cage-relative intermediate scattering function $F_s(\Delta t)$ defined in Eq.~\ref{eq:Fs}. The relaxation time $\tau$ is obtained from the time such that $F_s(\tau) = 1/e$ (horizontal dashed line). 
(a) $F_s(\Delta t)$ of the passive system at various densities. 
The blue dotted curve, which corresponds to $\phi = 0.91$, is a stretched exponential fit to obtain $\tau$ (at the red cross) if $F_s(\Delta t)$ does not reach $1/e$.
(b) $F_s(\Delta t)$ at $\phi = 0.91$ of different activity levels represented by P\'eclet number $\mathrm{Pe}$, and indicated by colors with the same legend in (c).
(c) Scaled relaxation time $\tau/\tau_0$ versus scaled density $\phi/\phi^*$ showing collapse on (Eq.~\ref{eq:VTF}, red line).
$\phi^*$ is obtained from the fit of systems able to reach a steady state (empty circles).
%
%
%where $\tau_0$ and $\phi^*$ is the relaxation time at the limit $\phi\rightarrow 0$ and ideal glass transition density, respectively. 
%$\phi^*$ is obtained from the density at which the VTF fit (Eq.~\ref{eq:VTF}, red line) diverges.
%In the passive system, we obtain $\phi^*(\mathrm{Pe}=0) = 0.93 \pm 0.005$.
%Note that the fitting is done only for densities that are completely equilibrated (empty circles), where $F_s(\Delta t)$ is a result of the time average.
The relaxation time of systems that do not reach a steady state (filled circles) is obtained from ensemble average of 50 simulations at the waiting time $t_\mathrm{w} = \num{1.5e4}$.
Uncertainty is smaller than the marker size.
}
\label{fig:supercooled}
\end{figure}

In general, the absolute motion of a particle in a glassy system can be split between the motion relative to the cage and the motion of the cage~\cite{mazoyer2009DynamicsParticlesCages}. In 2D, it is well known that the contribution of the motion of the cage cannot be neglected due to long wavelength sound modes~\cite{shiba2016UnveilingDimensionalityDependence}, also called Mermin–Wagner fluctuations~\cite{illing2017MerminWagnerFluctuations} which are system-size dependent. In order to focus on structural rearrangement and get rid of finite-size effects, we define the cage-relative motion of particle $i$ between time $t_0$ and time $t_1$,
\begin{equation}
\Delta \vect{r}^\mathrm{CR}_i(t_0,t_1) = \Delta \vect{r}_i(t_0,t_1) - \frac{1}{6}\sum_{j \in \mathrm{NN}_i} \Delta \vect{r}_j(t_0,t_1),
\end{equation}
where $\Delta \vect{r}_i(t_0,t_1)= \vect{r}_i(t_1) - \vect{r}_i(t_0)$ is the absolute displacement of particle $i$ and $\mathrm{NN}_i$ is the set of the six-nearest neighbors of particle $i$ at time $t_0$.

% Define Fs(q,t)
To quantify the system relaxation, we use the cage-relative self-intermediate scattering function $F\ls{s}(q,\Delta t)$, which is expressed as
\begin{eqnarray}
F\ls{s}(q, \Delta t, t\ls{w}) = \frac{1}{N} \left\langle \sum_{i = 1}^N \mathrm{e}^{\mathrm{i}\vect{q}\cdot
%[\vect{r}_i(t\ls{w} + \Delta t)-\vect{r}_i(t\ls{w})]
\Delta \vect{r}^\mathrm{CR}_i(t\ls{w}, t\ls{w} + \Delta t)
} \right \rangle,
\label{eq:Fs}
\end{eqnarray}
where $\langle\ldots\rangle$ denotes the average among different runs, either time average or ensemble average as described in Section~\ref{subsec:Sim_procedure}.
We choose $|\vect{q}| = 2\pi$, and we also average among four perpendicular directions in Fourier space.
For the sake of readability, in the following, we will use the simpler notation $F\ls{s}(\Delta t)$, and $t\ls{w}$ will be specified separately.

In Fig.~\ref{fig:supercooled}a, we show $F\ls{s}(\Delta t)$ obtained from the passive system, $\mathrm{Pe} = 0$, at various densities ranging from $\phi = 0.17$ to $1.03$.
For $\phi \leq 0.81$, $F\ls{s}(\Delta t)$ fully decays within the time interval of interest. This means that these systems have completely relaxed and we can perform a time-average from longer production runs.
As density increases, the relaxation proceeds in two steps with a well-defined plateau in between. The system takes longer time to relax and for  $\phi > 0.81$ $F\ls{s}(\Delta t)$ cannot fully decay within our measurement time. Such a system is not in a steady state and the results shown are obtained from the ensemble averaging at $t\ls{w}=\num{1.5e4}$.
The relaxation of absolute positions (not relative to the cage) display a slopped plateau due to long-range sound modes (not shown).
At very high densities, harmonic particles are known to exhibit counter intuitive behavior: due to the bounded potential particles can fully overlap~\cite{Jacquin2010}, thus relax faster than at more moderate densities~\cite{berthier2010IncreasingDensityMelts}. We confirm this reentrant behavior in our passive simulations for $\phi>1.1$ (not shown). In the following, we will remain at densities below this limit.

% Active systems
We show in Fig.~\ref{fig:supercooled}b how activity influences the system relaxation. At one fixed density $\phi = 0.91$, the passive system is not relaxed whereas the highly active system with $\mathrm{Pe} = 4.0$ manages to relax.
We can also see that the system relaxation is enhanced by activity: the higher the activity level, the faster relaxation.

% relaxation time
When $F\ls{s}(\Delta t)$ fully relaxes, we can define the relaxation time $\tau$ from the time such that $F\ls{s}(\Delta t)$ has reached the threshold $1/e$.
However, when we deal with correlation functions that do not completely relax, we cannot characterize the full relaxation and have to focus on the exit from the plateau.
For convenience, we fit this exit with a stretched exponential $A\exp\left(-(t/\mathcal{T})^\beta\right)$, where both $A=0.89$ and $\beta=0.45$ are fitted once at $\phi = 0.98$ in the passive case and then kept constant. $\mathcal{T}$ is thus the only fitting parameter.
In order to ensure a smooth connection between fully and poorly relaxing states, we intersect the stretched exponential resulting from the fit with the threshold $1/e$ to obtain a characteristic time $\tau$, as shown on Fig~\ref{fig:supercooled}a.
Although we will call this characteristic time a `relaxation time' it only quantifies the exit time from the plateau and not the full relaxation. In particular, we make no claim on the shape of $F\ls{s}(\Delta t)$ beyond our measurements.

% VTF form
Now that we have characterized the exit from the plateau with a single time scale, we can show its dependence on density in Fig.~\ref{fig:supercooled}c.
Starting from the lowest density, the relaxation time $\tau$ increases as the system is denser, and this is true for both passive and active systems.
The first part of the increase, is well-described by the Vogel-Tamman-Fulcher (VTF) form,
\begin{eqnarray}
\tau_\alpha = \tau_0\exp\left[\frac{B\phi^*}{\phi^* - \phi}\right],
\label{eq:VTF}
\end{eqnarray}
where $\tau_0$ is associated to the relaxation time obtained from the limit $\phi \rightarrow 0$, $B$ is a constant quantifying the steepness of $\tau(\phi)$ dependence, and $\phi^*$ is the density at which (Eq.~\ref{eq:VTF} diverges.
At each activity level we perform the VTF fit only using the densities that reach a steady state.
We fix the parameter $\tau_0$, which can be obtained directly from the simulation in the very dilute limit $\phi = 0.01$, and $B=0.75$ obtained from the passive case. 
This leaves $\phi^*$ as the only fitting parameter for the other activity levels.
Our results show that activity pushes glass transition $\phi^*$ towards a higher density (inset) as also reported in previous simulation~\cite{Ni2013} and experimental~\cite{Klongvessa2019_PRE} works.
% note on B
We note that in Ref.~\onlinecite{Ni2013} the value of $B$ was reported to be decreasing with $\mathrm{Pe}$. However, in this work, we are focusing on the very low activity limit where this decrease is negligible. Even if we leave $B$ as a free parameter, the difference in $B$ between the passive case and the highest activity level of interest ($\mathrm{Pe} = 4$) is still smaller than the uncertainty from the fitting, which is about 0.1.

The system relaxation is well-described by the VTF relation for $\phi/\phi^*(\mathrm{Pe}) < 0.85$, at which the system is in an ergodic supercooled state.
As it has been shown in the experimental work~\cite{Klongvessa2019_PRE}, our simulation verifies that the density-dependence relaxation of the passive and active systems in the supercooled state can be collapsed into one master curve. That is to say, the mapping between the passive (equilibrium) and active (nonequilibrium) system in the supercooled state is valid.

% ===============================================================

\subsection{Dynamics in the glass state}

% beyond the glass transition
At high densities, we observe on Fig~\ref{fig:supercooled}a that the increase of relaxation time $\tau$ is slower than the VTF trend and we notice the beginning of a saturation of $\tau$ at $\phi/\phi^* \approx 1$, which is typical of soft particles\cite{philippeGlassTransitionSoft2018} and in the case of bounded potential is the signature of a reentrant melting~\cite{berthier2010IncreasingDensityMelts}.
Although in the passive case the maximum of $\tau(\phi)$ is located at a higher density than all explored densities, at $\mathrm{Pe}\leq 2$ we observe this maximum within our explored density range. The position of the maximum shifts toward lower densities with activity. This behavior is reminiscent of the temperature response of harmonic spheres~\cite{Jacquin2010}.
However, the previous mapping between passive and active states fails in the glass state: for increasing activity the deviations from VTF occurs at lower reduced density and the curves do not collapse. 
In other words, beyond the glass transition the behavior of the active system cannot be described by an effective passive system.
We actually observe that the maximum of $\tau(\phi)/\tau_0$ exhibits a nonmonotonic behavior with the activity level. All curves at $0.2\leq\mathrm{Pe}\leq 1$ are higher than the passive case, whereas all curves at $\mathrm{Pe}>1$ are lower than the passive case.
Although a shift in reentrant melting might explain the decrease of rescaled relaxation time at high $\mathrm{Pe}$, such a monotonic shift cannot be the cause of the slowing down at low $\mathrm{Pe}$.
%in an active glass has been observed as well from the experimental system of Janus particles~\cite{Klongvessa2019_PRL}.
%In the following, we shall focus on the glassy nonergodic regime and investigate further its relaxation.

\begin{figure}
\includegraphics{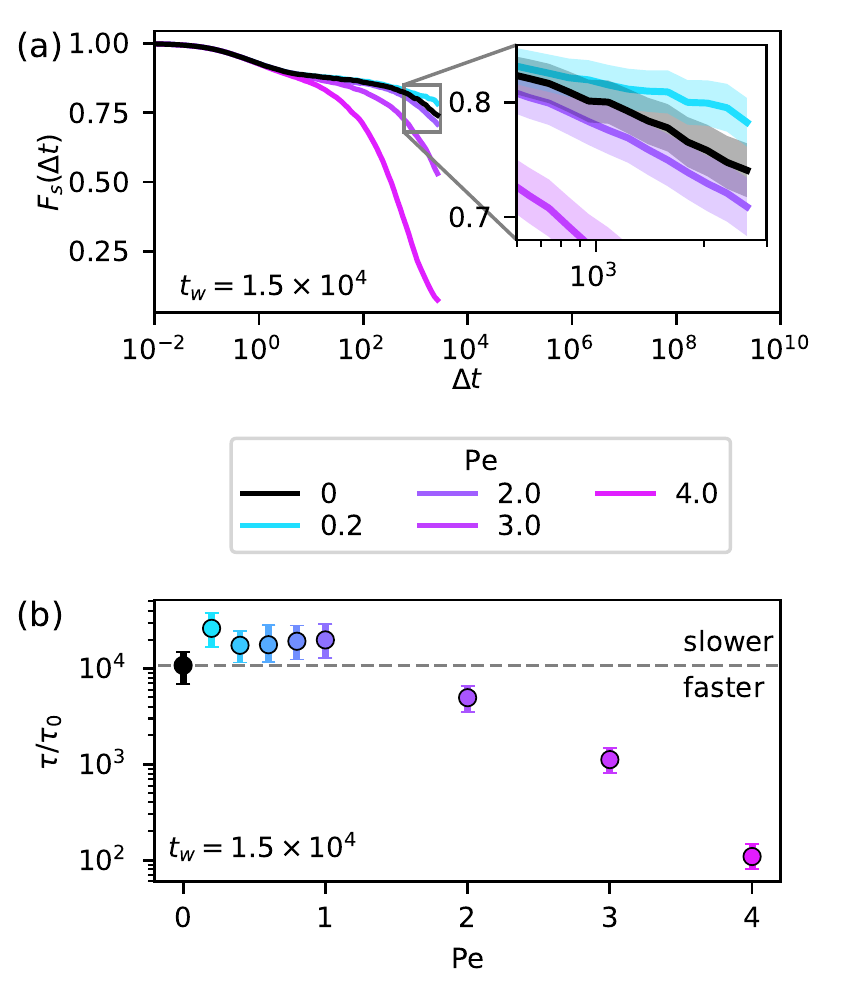}%
\caption{
(a) Self-part of the cage-relative intermediate scattering function $F_s(\Delta t)$ for various activity levels $\mathrm{Pe}$ at $\phi =1.03$ for the waiting time $t_\mathrm{w} = \num{1.5e4}$. 
(Inset) Highlight of the nonmonotonic behavior of relaxation time $\tau$ with the activity level.
(b) Activity dependence of $\tau/\tau_0$ showing the nonmonotonic behavior.
}%
\label{fig:Fs_nonmono}
\end{figure}

% Fs 
To confirm that the increase in relaxation time is not an artifact of our fitting procedure, we display directly the dependence of $F\ls{s}(\Delta t)$ on activity level at high density ($\phi=1.03$) in Fig.~\ref{fig:Fs_nonmono}a.
In particular, we focus on the exit from the plateau. The inset of Fig.~\ref{fig:Fs_nonmono}a displays error bars on $F\ls{s}(\Delta t)$ obtained by ensemble averaging over $n=50$ independent trajectories: $\mathrm{Er}\left[F\ls{s}(\Delta t)\right] = \mathrm{std}_n\left[F\ls{s}(\Delta t)\right]/\sqrt{n}$.
We observe that at $\mathrm{Pe}=0.2$ (cyan curve), $F\ls{s}(\Delta t)$ is significantly above the passive case (black curve), indicating a delayed exit from the plateau.
However, when activity is high enough ($\mathrm{Pe} \geq 2$, purple to magenta curves), the exit from the plateau is faster than the reference passive case and the system relaxes all the more fast than $\mathrm{Pe}$ increases.

% relaxation time
In Fig.~\ref{fig:Fs_nonmono}b, we quantify the system relaxation using the relaxation time $\tau$ extracted from $F\ls{s}(\Delta t)$, normalized by its low density value $\tau_0(\mathrm{Pe})$.
Even once the trivial enhancement of the dynamics by self-propulsion taken into account, the relaxation time drops by two orders of magnitude between the passive ($\mathrm{Pe}=0$) and the most active system ($\mathrm{Pe}=4$).
However at low activities ($0<\mathrm{Pe} \leq 1$) the relaxation time is slower than the passive case by a factor $\approx 2$. Qualitatively similar results are obtained without the normalization by $\tau_0$ (not shown).

This nonmonotonic behavior is the key observation of our previous experiments~\cite{Klongvessa2019_PRL},
and we show here that it can be recovered within a simple numerical glass model.
Beyond this first achievement, we describe in the next section the numerical phase diagram to demonstrate that the in silico system is fully consistent with experiments to date.
Then we are in a position for carrying new detailed characterizations aiming at deciphering the origins of this nonmonotonic behavior.
%\del{As we have managed to reproduce this phenomenology in silico, we shall now characterize it further and explore the minimal set of conditions to observe it.}

% The different behavior between the shorter and the longer waiting time provides us a fingerprint that aging could be a crucial condition for the nonmonotonic behavior.
% This leads to another question of how activity influences aging.
% We will address this point again in Section~\ref{subsec:aging}.
% But first, we shall take a wider look at the activity-dependent relaxation of the early age system at various densities. The aim of the next part is to locate the nonmonotonic behavior in a phase diagram.

% ==================================================

\subsection{State diagram}

\begin{figure}
    \centering
    \includegraphics[width = 0.5\textwidth]{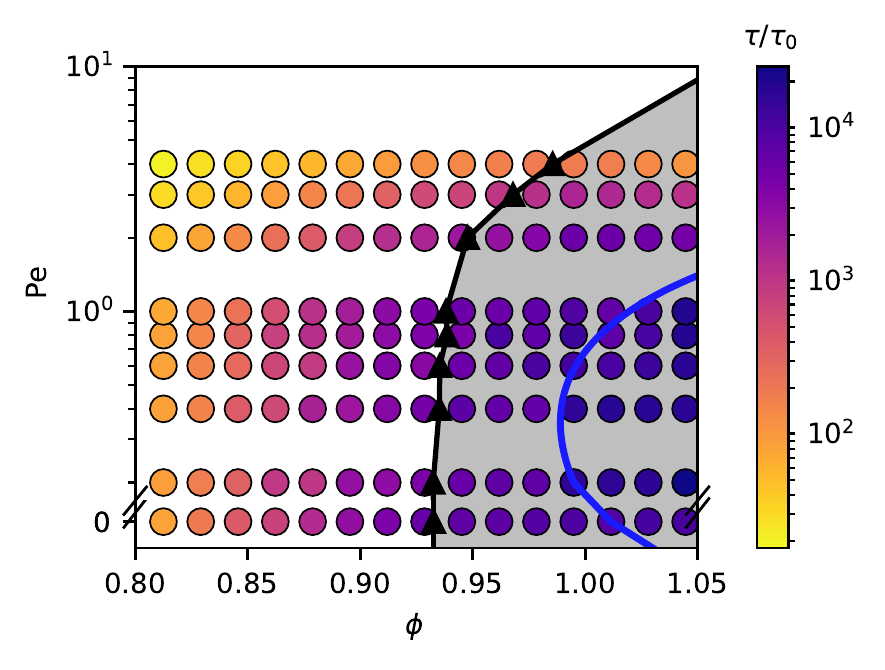}
    \caption{Activity level $\mathrm{Pe}$ and density $\phi$ state diagram showing the dependence of scaled relaxation time $\tau/\tau_0$, which is shown by colored circles.
    The triangles represent the activity-dependent glass transition density $\phi^*(\mathrm{Pe})$.
    Inside the glass state, which is displayed by the grey area, the blue curve is a guide for the eye of a contour line at which $\tau/\tau_0 \approx \num{9e3}$, that corresponds to the saturation level of $\tau$ in the passive system.
    %$\tau \approx 4t\ls{w}=\num{6e4}$
    The results are obtained at the waiting time $t\ls{w} = \num{1.5e4}$.}
    \label{fig:phase_diagram}
\end{figure}

The system behavior, which depends on both density $\phi$ and activity level $\mathrm{Pe}$, is summarised in the %$\mathrm{Pe} - \phi$ 
state diagram shown in Fig.~\ref{fig:phase_diagram}.
The rescaled relaxation time $\tau/\tau_0$ is represented by colors.
The fitted divergence $\phi^*(\mathrm{Pe})$ of (Eq.~\ref{eq:VTF}) are displayed by the black triangles, and we can use them to illustrate the glass region (grey zone).
In the passive system, the saturation of relaxation time occurs around $\tau/\tau_0 \approx \num{9e3}$.
We draw a contour line (blue line) around the region exceeding that value.

The system behavior at one fixed density can be understood such that we travel vertically from zero to nonzero $\mathrm{Pe}$ on the state diagram.
%
% nonmono here
Within this visualization, we can locate the nonmonotonic behavior (the rise and fall of $\tau$) occurs at sufficiently high density inside the glass state.
Starting from the passive state at about $\phi>0.98$, $\tau$  increases when the system becomes slightly active, and stays higher than the passive $\tau$ at low level of activity ($0<\mathrm{Pe}\leq 1$). As the activity level increases further, $\tau$ drops and we can see that the system is about to leave the glass state in the upwards direction.

%This phase diagram is a support to Fig.2b in Ref.~\onlinecite{Klongvessa2019_PRL}, which came from the experimental results.
This state diagram is consistent with our experimental measurements~\cite{Klongvessa2019_PRL}.
It shows that the behavior of an active glass at short aging time cannot be mapped on the passive soft sphere state diagram simply by replacing the temperature by $\mathrm{Pe}$.
%It shows that the conventional glass transition phase diagram for a passive system, where $\mathrm{Pe}$ is replaced by temperature, is not sufficient to describe an active glass system.
%In other words, the role of activity in a form of self-propulsion cannot simply interchange with effective temperature for the glass phase.

% At this point, we can summarise the system behavior at different values of model parameters,
% However, the remaining question is the origin of this behavior.
% In the next section, we shall look at the microscopic behavior in the glassy state, and characterize the relaxation mechanisms unique to active matter.
%system dynamics with a length scale.

%\del{At this point, we have successfully reproduced with our simulations the nonmonotonic phenomenology observed experimentally.
%In the following section, 
%}
Now that our simulations proved to capture the salient features of experimental active colloidal glasses,
we use them to zoom in to the cage level in order to investigate the influence of activity on structural relaxation behavior.

\section{Cage exploration and escape}
\label{sec:cage_explore}

\begin{figure}
\includegraphics{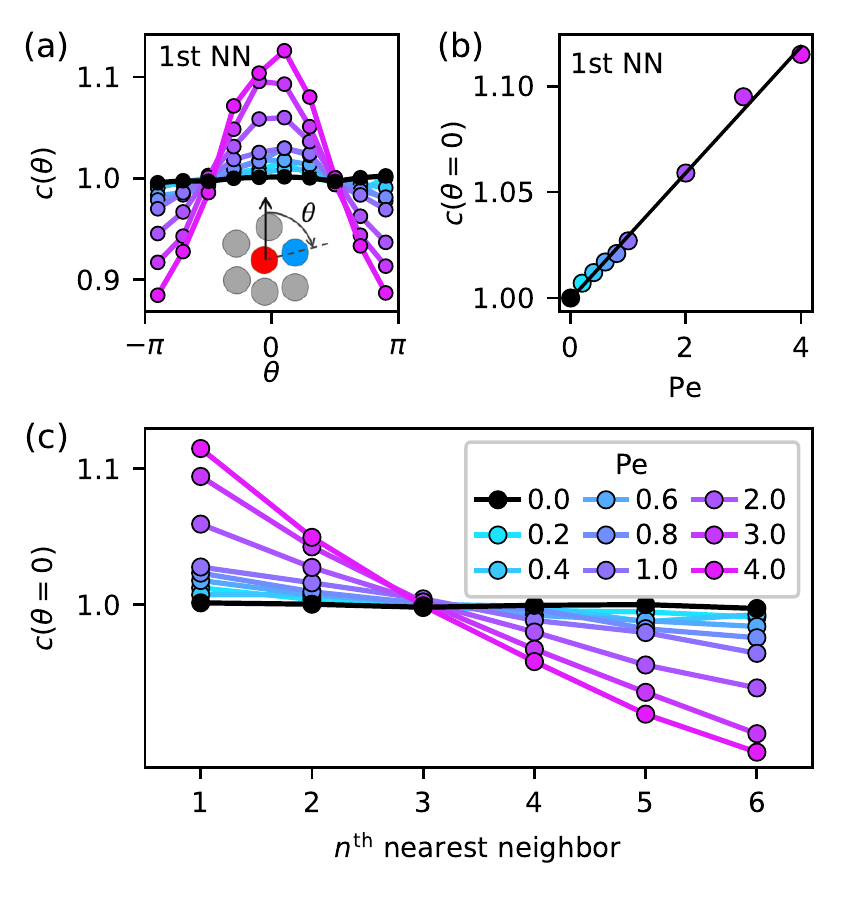}%
\caption{
(a) Azimuthal correlation function $c(\theta)$ between a particle and its first nearest neighbor with respect to particle's orientation. Activity levels $\mathrm{Pe}$ are shown by colors in the same representation as in (c).
(Inset) A sketch illustrates how $c(\theta)$ is defined for the first nearest neighbors.
(b) $c(\theta = 0)$ calculated from the first nearest neighbor as a function of activity levels $\mathrm{Pe}$.
The black line is a linear fit.
(c) $c(\theta = 0)$ as a function of the selected nearest neighbor, from the first to the sixth nearest, at various activity levels $\mathrm{Pe}$.
Uncertainty is smaller than the marker size.}
\label{fig:NN_ang_dist}
\end{figure}

%re-explain DEAD model and what we should observe
Previously, we have explained the slowing down of structural rearrangements at low activities by an inefficient exploration of the cage~\cite{Klongvessa2019_PRL}.
%To sum up briefly this argument, we considered that a small amount of self propulsion does not affect the height of energy barriers hindering rearrangements.
%However, the nature of the motion within the cage affects the frequency at which the particle could attempt to hop out of its cage. A Brownian particle randomly explores its cage and can often test weak points of the cage (low energy barriers). A particle with persistent directed motion will be able to test a single direction until it reorients. At the scale of the cage, persistent motion can be less efficient than Brownian motion, thus the attempt frequency lower and the structural rearrangement at low activities slower than in the equivalent passive system.
%
With the present simulations, we have the opportunity to test the Deadlock by the Emergence of Directionality (DEAD) model. Contrary to experiments, we have access to the orientation of the propulsion of the particles $\theta_i$ and thus can test our hypothesis on cage exploration.

\subsection{Angular distribution of nearest neighbors}

In our previous attempt to understand the experimental non monotonic behavior, we argued that the DEAD occurs when a particle is more confined by its self-propulsion than by the cage~\cite{Klongvessa2019_PRL}. 
Indeed, for times shorter than $\tau_\mathrm{R}$, the propulsion force can be considered of constant direction and in analogy with sedimentation-diffusion, the confinement length $\lambda_\mathrm{P}=\mathrm{Pe}^{-1}\sigma$ emerges. For Peclet numbers larger than 1, a particle against a hard barrier is confined within less than its own diameter. 
Here, we will check whether this confinement effect exists in our simulations.
%However in our simulations we always have $\lambda_\mathrm{P}>a$.
    
% define c(\theta)
% Until now, we only characterised the system using solely particle coordinate ($x_i,y_i$). 
% Actually, another information that we can obtain directly from the simulation is particle orientation ($\theta_i$).
For an active particle ($\mathrm{Pe}>0$), $\theta_i$ is the direction of self-propulsion. For a passive particle ($\mathrm{Pe} = 0$), we simulated the evolution of $\theta_i$ although the force attached to the particle is null.
Here we investigate how the nearest neighbors of particle $i$ distribute around it with respect to $\theta_i$.
The azimuthal correlation function $c(\theta)$ is defined as,
\begin{eqnarray}
c(\theta) = \frac{2\pi}{\Delta\theta}\frac{1}{N} \frac{1}{n} \left\langle \sum_{i = 1}^N \sum_{j\in \mathrm{NN}_i} \delta(\theta_{ij}) \right\rangle,
\label{eq:azimuthal_corr}
\end{eqnarray}
where $\Delta\theta$ is the bin size.
$\mathrm{NN}_i$ is a set containing $n$ nearest neighbors of particle $i$.
$\theta_{ij}$ is the angular position of particle $j$ with respect to the orientation of particle $i$,
\begin{eqnarray}
\theta_{ij} = \theta_i - \tan^{-1}\left(\frac{y_j - y_i}{x_j - x_i}\right).
\end{eqnarray}
$\delta$ in Eq.~\ref{eq:azimuthal_corr} is an rectangular impulse function such that
\begin{equation}
\delta(\theta_{ij}) = 
\begin{cases}
1, &\text{if $\theta < \theta_{ij} < \theta + \Delta\theta$} \\
0, &\text{otherwise.}
\end{cases}
\end{equation}
%where $\langle \dots \rangle_i$ is the average among all particles. $\theta_{ij}$ is the orientation of the vector pointing from particle $i$ to particle $j$. 
%$\theta_i - \theta_{ij}$ is therefore the angular position of particle $j$ with respect to the orientation of particle $i$.
%$\mathrm{NN}_i$ is a set of nearest neighbors of particle $i$ that we will vary in the following. 

% c(\theta) 1st NN
First, let us focus on the first nearest neighbor.
In Fig.~\ref{fig:NN_ang_dist}a, we show the azimuthal correlation $c(\theta)$ of the nearest neighbor.
The inset illustrates the notation of the direction $\theta$:
$\theta = 0$ corresponds to the neighbor being `in front' of the particle, i.e. in the direction of self-propulsion for an active particle (arrow). Conversely, the `rear' corresponds to $\theta = \pm \pi$.
For a passive particle, the orientation does not mean anything since there is no self-propulsion. Consistently, $c(\theta)$ displays uniform distribution (black line).
This means that the probability of finding the nearest neighbor is equal for all directions.

In active systems, the distribution becomes nonuniform.
There is a higher chance of finding the nearest neighbor at the front of the particle and lower chance at the rear.  
The degree of nonuniformity is stronger with the increasing activity level as the peak of $c(\theta)$ increases linearly with $\mathrm{Pe}$, as shown on Fig.~\ref{fig:NN_ang_dist}b. 
This nonuniformity indicates that the closest neighbor mostly locates at the front of an active particle.
In other words, an active particle is pushing its cage due to the self-propulsion.

In Fig.~\ref{fig:NN_ang_dist}c we show the value of the peak of the distribution in front of the particle $c(\theta = 0)$, function of the selected neighbor, from the first to the sixth nearest.
We confirm the effect of self-propulsion, with active particle pushing toward its nearest neighbor and away from the rear of the cage.

This observation is consistent with the hypothesis of our DEAD model. In the following, we will explore whether the escape from the cage is at the origin of the observed nonmonotonic response.

\begin{figure}
\includegraphics{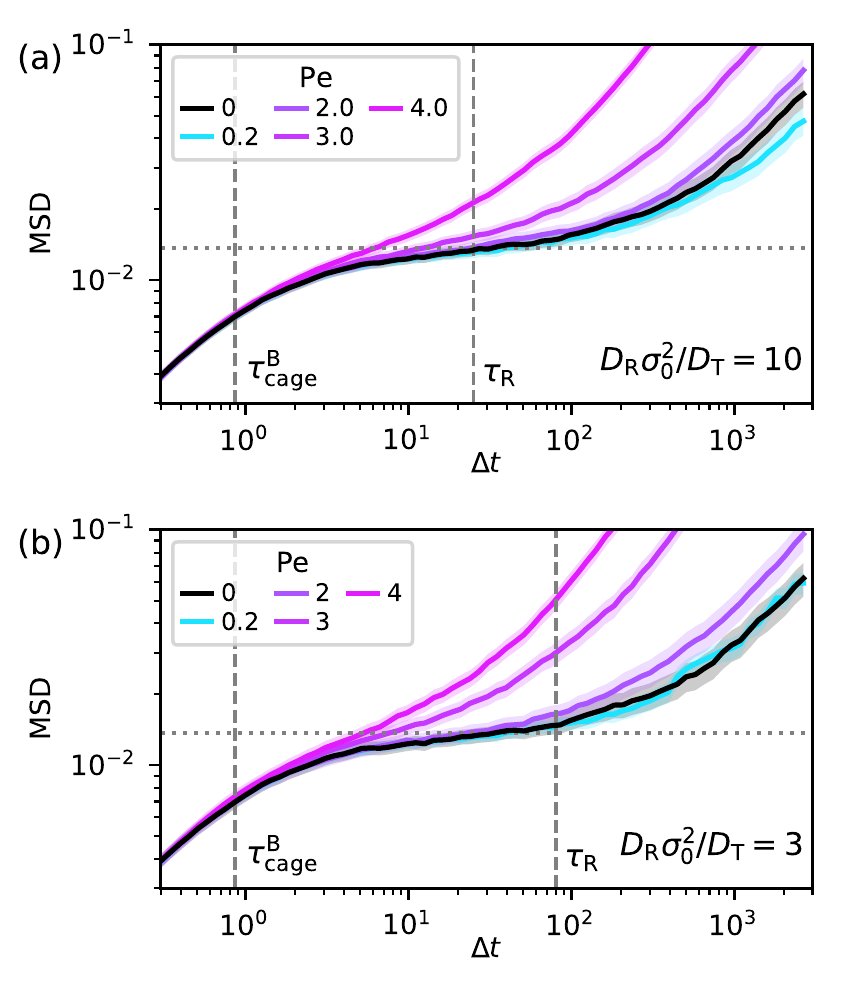}%
\caption{
Cage-relative mean square displacement $\mathrm{MSD}$ at $\phi = 1.03$ and various activity levels $\mathrm{Pe}$.
Colored area indicate the error bars from ensemble averaging.
Horizontal dotted line indicates the squared cage size $a^2$.
Vertical dashed lines indicate respectively the time to explore the cage by Brownian motion $\tau\ls{cage}^\mathrm{B}$ and the rotational time $\tau_\mathrm{R}$.
(a) For a rotational diffusion set by $\sigma^2 D_\mathrm{R}/D_\mathrm{T}=10$, consistently with other figures.
(b) For a longer rotational time such that $\sigma^2 D_\mathrm{R}/D_\mathrm{T}=3$, where we notice that the slowing down at low Peclet is less pronounced. 
}
\label{fig:cage_motion}
\end{figure}

\subsection{Length and timescales of cage exploration}
\label{subsec:length_and_time_scale}

To quantify the length and time scales of cage exploration, we define the cage-relative mean square displacement:
\begin{equation}
\mathrm{MSD}(\Delta t, t\ls{w}) =  \left\langle\frac{1}{N}\sum_{i=1}^N|\Delta \vect{r}^\mathrm{CR}_i(t\ls{w}+\Delta t, t\ls{w})|^2 \right\rangle,
\label{eq:CR_MSD}
\end{equation}
where $t\ls{w}=\num{1.5e4}$. Thus we immediately simplify the notation to $\mathrm{MSD}(\Delta t)$.

In Fig.~\ref{fig:cage_motion}a, we show the cage relative mean square displacement at $\phi = 1.03$ for various activity levels.
The passive system (black line) shows the typical behavior of glassy systems: short time diffusion inside the cage, long time diffusion from cage to cage, and a plateau at intermediate times $10\lesssim \Delta t \lesssim 10^2$.
If $a$ is the typical cage size, the height of this plateau is $a^2$. 
We can thus estimate $a \approx 0.125$.
The typical time to diffuse across the cage for a passive particle is $\tau\ls{cage}^\mathrm{B} = a^2/4D\ls{T} \approx 1$.

On Fig.~\ref{fig:cage_motion}a we observe a non monotonic response of the length of the plateau to activity, consistent with Fig.~\ref{fig:Fs_nonmono}a. What the MSD teaches us, is the details of the particle behavior within the cage and at very short times.
Low levels of activity do not affect the height of the plateau and thus the size of the cage. 
Furthermore the short times behaviors are also superimposed, meaning that the nature of the displacement within the cage is not significantly altered by the weak self-propulsion force.
This observation goes against the central hypothesis of the DEAD model, i.e. inefficient cage exploration by directed motion.

The DEAD model supposes than the size of the cage is shorter that the persistence length, a condition equivalent to
\begin{equation}
    %\ell_\mathrm{P} > a \Leftrightarrow 
    \mathrm{Pe} > a \frac{\sigma^2 D_\mathrm{R}}{D_\mathrm{T}}.
    \label{eq:persistence_cage}
\end{equation}
With our simulation parameter $\sigma^2 D_\mathrm{R}/D_\mathrm{T}=10$ and at $\phi=1.03$, we expect the DEAD behavior only for $\mathrm{Pe}>1.25$.
Instead, we observe it for $\mathrm{Pe}<1$ see Fig.~\ref{fig:Fs_nonmono}b.
Furthermore, (Eq.~\ref{eq:persistence_cage}) predicts that the effect would be visible for a larger range of Peclet number if $D_\mathrm{R}$ decreases, i.e. the rotational time increases.
Consistently, the DEAD model predicts that the magnitude of the slowdown should be $\tau_\mathrm{R}/\tau\ls{cage}^\mathrm{B}$ and thus increases with the rotational time.
Contrary to both predictions, the nonmonotonic behavior is less visible when $\sigma^2 D_\mathrm{R}/D_\mathrm{T}=3$, see Fig.~\ref{fig:cage_motion}b where the passive and lowest activity MSD are within the error bars of each other.

%\del{We have thus demonstrated that the predictions of the DEAD model do not apply to the nonmonotonic behavior we observe in our simulations.}
Within our simulations, we have thus demonstrated that cage exploration is indeed skewed by activity. However unlike our initial DEAD proposal, 
the slowdown is more pronounced for shorter rotational time and shorter persistent length, and the mean dynamics within the cage seems to be little affected by the persistent self-propulsion.
Therefore, DEAD does not properly account for the nonmonotonic response with activity, and we need to look for %\del{an other explanation of the nonmonotonic response to activity.}
alternative mechanisms.
In the following section, we address the link between this response and the hallmark of nonergodicity: aging.

\section{Aging}
\label{sec:aging}

\begin{figure}
\includegraphics{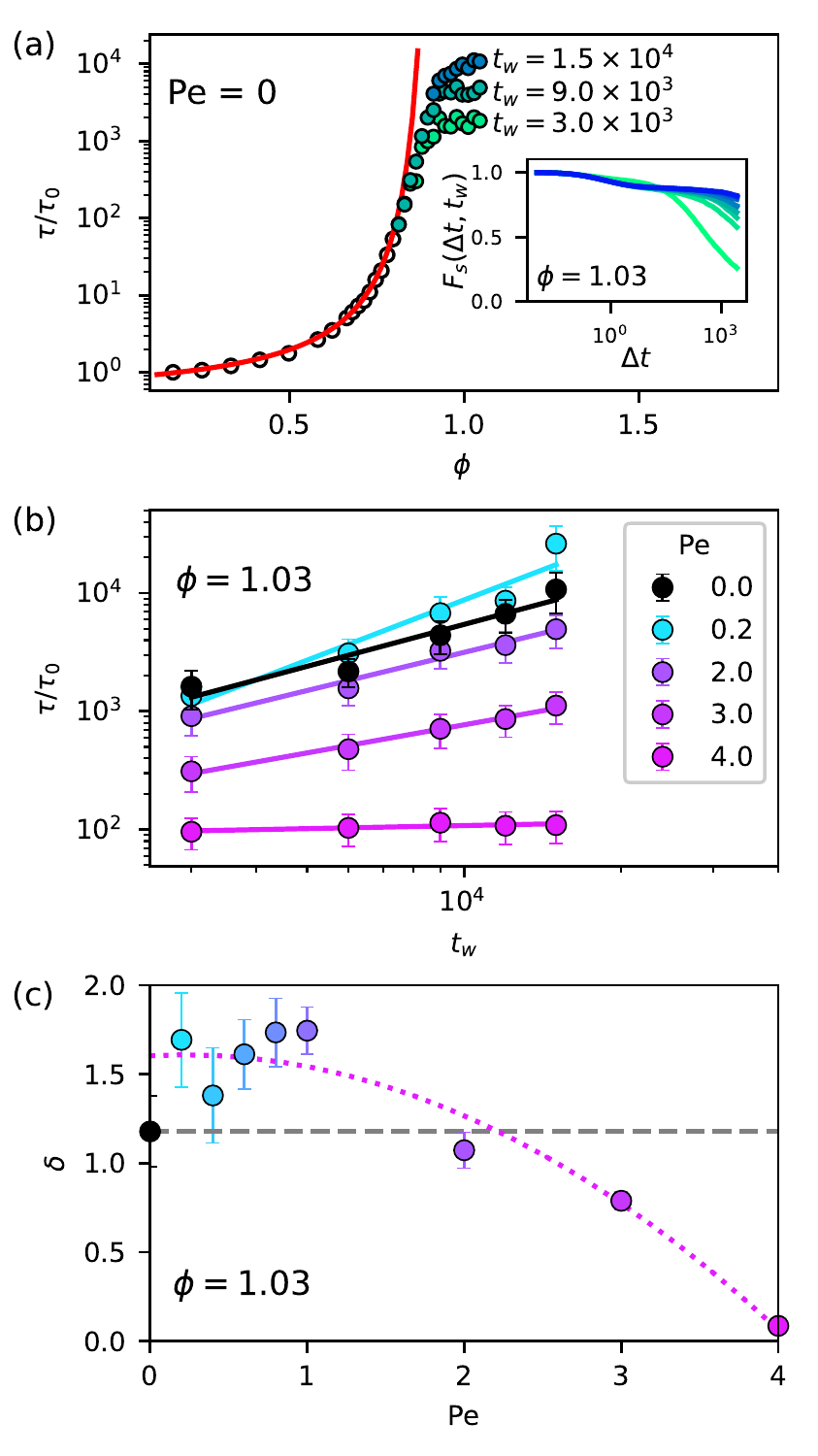}%
\caption{
(a) Density dependence of relaxation time $\tau$ of the passive system ($\mathrm{Pe} = 0$). The empty circles are steady state systems and are used for the fit of VTF relation (red line).
The waiting time dependence of $\tau$ is illustrated by the filled circles colored by waiting time $t_\mathrm{w}$. %{\color{red}The transparent area show standard error of the ensemble average CANNOT SEE WHAT IT MEANS IN FIGURE}. 
(inset) $F_s(\Delta t)$ at $\phi = 1.03$ colored by $t_\mathrm{w}$ from green (\num{3e3}) to blue (\num{5.7e4}).
(b) Scaled relaxation time $\tau/\tau_0$ as a function of waiting time $t_\mathrm{w}$ at $\phi = 1.03$ of various activity levels $\mathrm{Pe}$.
The solid lines are the power law fit $\tau/\tau_0 \sim t_\mathrm{w}^\delta$, where $\delta$ is an aging exponent.
(c) Aging exponent $\delta$ function of activity level $\mathrm{Pe}$.
The dotted line is a parabolic function eye-guide (see main text) and the dashed line represents the value of $\delta$ in the passive case.
}%
\label{fig:tw_dependece}
\end{figure}

The state of nonergodic systems depends on their history. 
In particular, the relaxation dynamics of a system brought to a glass state slows down with the time since preparation, a phenomenology called aging. 
In the inset of Fig.~\ref{fig:tw_dependece}a we confirm that phenomenology in our passive system by plotting $F_s(\Delta t)$ for various waiting times ranging from $t\ls{w} = \num{3.0e3}$ (green) to $\num{5.7e4}$ (blue) at one fixed density $\phi = 1.03$.
As $t\ls{w}$ increases, the exit from the plateau is more and more delayed.
For the longest waiting times $t\ls{w}> \num{1.5e4}$, the exit time from the plateau is not reliably captured.
However, for each waiting time $t\ls{w}\leq \num{1.5e4}$, we can fit the exit from the plateau as in Sec.~\ref{subsec:results_supercooled} to define a characteristic time $\tau(t\ls{w})$.
This procedure can be generalized at various densities, as shown in Fig.~\ref{fig:tw_dependece}a. At low densities, all waiting times follow the same VTF relation between $\phi$ and $\tau$. However in the nonergodic regime, the saturation level of the relaxation time occurs at lower values in younger systems (shorter $t\ls{w}$).

%Aging of passive systems beyond the glass transition is well characterized~\cite{struik1977physical,berthier_theo_2011}. For instance, the relaxation time $\tau$ increases as a power-law of the waiting time $t\ls{w}$.

In the following, we study how the activity-dependence of the relaxation depends on the preparation history, and how this aging relates to the observed nonmonotonic response of the relaxation to activity.

%fig{fig:tw_dependece}a 
%aging in the passive case
%In our system, for $\phi>\phi^*$ we expect that the relaxation depends on the waiting time. 
%Up to this point, all of the nonequilibrated results were obtained at the fixed waiting time $t\ls{w} = \num{1.5e4}$.
%In Fig.~\ref{fig:tw_dependece}a, we show the relaxation time $\tau$ as a function of density $\phi$ obtained from the passive system at various waiting times ranging from $t\ls{w} = \num{3.0e3}$ to $\num{5.7e4}$.
%It can be seen that the younger systems (shorter $t\ls{w}$) have more deviation from the VTF relation than the older systems (longer $t\ls{w}$).
%Besides, at high densities $\phi > 0.9$, the younger systems display the saturation of $\tau$, while $\tau$ continues increasing in the older ones.
%We can also compare the system relaxation at one fixed density $\phi = 1.03$ but different waiting time as shown in the inset: the longer the waiting time, the slower the relaxation, which is the classical aging phenomenon.
%In the following, we will see how activity influences this scenario.

\subsection{Aging in the active system}

% compare tau-t_w in diff Pe
%It is well-known for a passive system that relaxation time $\tau$ increases with waiting time $t\ls{w}$ and has a power-law relation~\cite{struik1977physical,berthier_theo_2011}.
We show in Fig.~\ref{fig:tw_dependece}b the dependence of $\tau$ on $t\ls{w}\leq \num{1.5e4}$ comparing between different activity levels $\mathrm{Pe}$ at the same density $\phi = 1.03$.
As expected from previous studies on passive glassy systems~\cite{struik1977physical,berthier_theo_2011}, we observe at $\mathrm{Pe} = 0$ a power-law increase of $\tau \propto t\ls{w}^\delta$, where $\delta$ is the aging exponent.
By contrast for $\mathrm{Pe} = 4.0$, there is no dependence on waiting time. At this high level of activity aging effects disappear, the system is ergodic even at $t\ls{w} = \num{3.0e3}$, as self-propulsion promotes system relaxation.
For intermediate activity levels, we observe power-law aging of varying exponents and prefactors.
On Fig.~\ref{fig:tw_dependece}c we show the dependence of the aging exponent on $\mathrm{Pe}$.
Overall, the aging exponent drops from $\delta \approx 1.6$ ($\mathrm{Pe}<1$) to $\delta \approx 0 $ ($\mathrm{Pe} = 4$). This seems consistent with recent studies on thermal~\cite{janzen2021agingthermalglass} and athermal\cite{Mandal_Sollich_PRL2020_multi_aging} active glasses.
In particular \textcite{Mandal_Sollich_PRL2020_multi_aging} have described the evolution of the aging exponent as a parabola in an athermal active glass in a regime where the persistence time is very large.
Although the persistence time in our system is small ($\tau\ls{R} = 25$) with respect to relaxation time, we nevertheless observe a decrease of $\delta$ compatible with a parabola for $\mathrm{Pe}>1$ (dotted line on Fig.~\ref{fig:tw_dependece}c).

The effect of low amount of activity ($\mathrm{Pe}<1$) on aging is more complex. 
In particular, we observe that $\delta$ at $\mathrm{Pe}=0$ lies significantly below the parabola, close to $\delta=1$ often observed in passive systems~\cite{arceriGlassesAgingStatistical2020}. 
It means that the passive system ages slower than slightly active systems.
As a consequence, we can observe a crossing on Fig.~\ref{fig:tw_dependece}b: 
at short waiting times $t\ls{w} = \num{3.0e3}$, relaxation times monotonically follow $\mathrm{Pe}$, however at longer waiting times the lowest activities can exhibit longer relaxation times than the passive system.
From this, we can deduce that the nonmonotonic response of the relaxation develops with waiting time and is an effect of activity-enhanced aging.

We note that the nonmonotonic behavior we investigate here has been observed only with soft interactions (the Yukawa potential of our past experiments and our present harmonic potential simulations). 
Our results link this behavior with the observation of a saturation or a maximum in $\tau(\phi)$, which level depends on the age of the system, see Fig.~\ref{fig:tw_dependece}a.
%Such saturation has been observed in experimental soft systems~\cite{philippeGlassTransitionSoft2018}, but better designed numerical protocols are needed to explore it systematically.

Overall, the effect of activity appears quite reminiscent of phenomena associated with passive glasses under mechanical solicitation. There indeed, small shear can trigger a glass over-aging while higher ones can induce its partial to complete system rejuvenation~\cite{Viasnoff.2002}. It is remarkable to note that the same qualitative scenario occurs here from a local energy injection by active particles. 
%The increased complexity of the response, with the cross-over from activity-induced over-aging to rejuvenation being age-dependent might be related to the fact that solicitation is more of a fixed (active) force form rather than of a fixed displacement.
%{\color{red} Check what you think of last more hypothetical sentence. I felt that with an aging sample fixed stress and fixed strain might not yield identical response. Note that it could be also wise not to try to map all of active system to passive one to put forward their singularity...}
Shear-enhanced aging is often understood in terms of potential energy landscape: mechanical energy helps the young system to hop out of relatively shallow energy minima into deeper ones, increasing its age~\cite{Viasnoff.2002}.
If this picture applies to our system, we expect activity-enhanced aging to be more pronounced for systems that are prepared in even more shallow minima.
We will test this hypothesis below.

\subsection{Protocol dependence}
\label{sec:protocol_dependence}

\begin{figure}
    \centering
    \includegraphics{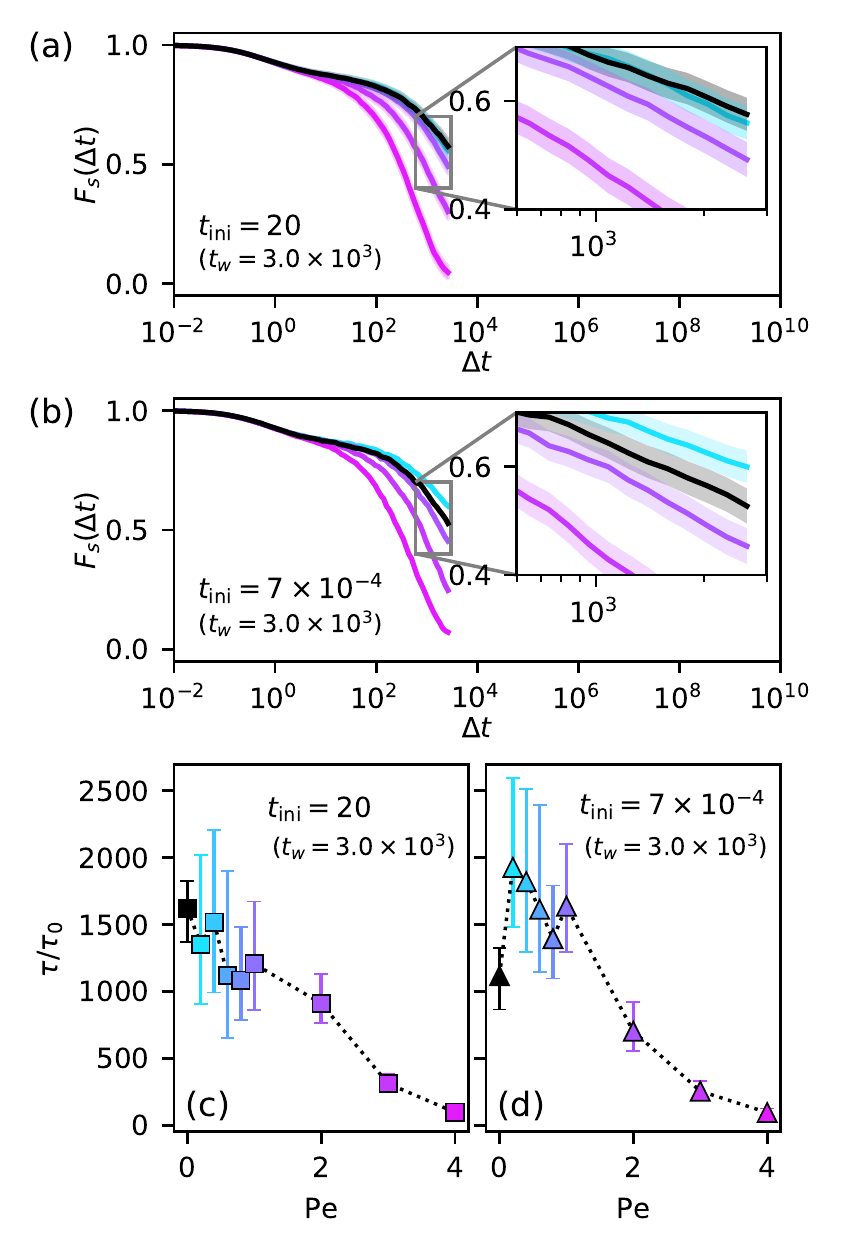}
    \caption{Dependence on the initiation time.
    (a,b) Self-intermediate scattering function $F\ls{s}(q,\Delta t)$ at $\phi = 1.03$ and waiting time $t_\mathrm{w} = \num{3e3}$ for 
    (a) $t_\mathrm{ini} = 20$, 
    and (b) $t_\mathrm{ini} = \num{7e-4}$.
    (c-d) Activity dependence of $\tau$ comparing between different initialization time: $t_\mathrm{ini} = 20$ (c), and $t_\mathrm{ini} = \num{7e-4}$ (d).}
    \label{fig:newrun}
\end{figure}
    
Since aging is by nature protocol dependent, we also checked the relevance of the initialization stage of our simulations. Indeed until now, as explained in Section~\ref{sec:method}, we prepare our system by letting a random configuration of harmonic spheres evolve through passive dynamics during $t\ls{ini} = 20$ in order to separate overlapping particles.
During $t\ls{ini}$ the potential energy decays by $\Delta E_\mathrm{p}(\phi)$.
We now introduce a new protocol where we shorten $t\ls{ini}$ to \num{7e-4} that correspond in average to a drop of potential energy of only $\Delta E_\mathrm{p}(\phi)/4$ during initialization at $\phi = 1.03$.
This new protocol produces configurations that are less relaxed, at which we set $t\ls{w}=0$ and turn on the self-propulsion.

In Fig.~\ref{fig:newrun} we compare the response to activity of relaxation between these two preparation protocols differing only in the duration of the initialization stage.
Fig.~\ref{fig:newrun}a and b display $F\ls{s}(q,\Delta t)$ at $t\ls{w} = \num{3e3}$. For this relatively short waiting time, we have observed until now ($t\ls{ini} = 20$) that the exit from the plateau become monotonically earlier with activity, as confirmed in Fig.~\ref{fig:newrun}a.
By contrast, at the same conditions but with a shorter initialisation time $t\ls{ini}=\num{7e-4}$, the exit time from the plateau responds nonmonotonically to activity, as shown in Fig.~\ref{fig:newrun}b.
The response of relaxation times to activity confirm the difference between the two preparation protocols, as show in Fig.~\ref{fig:newrun}c and d.
Therefore, starting from a poorly relaxed configuration allows to observe the nonmonotonic response of the relaxation in younger systems.

With the new protocol, we have confirmed that the nonmonotonic response was intimately related to nonergodic phenomenology. In particular, a significant amount of aging within the observation window is necessary, which can be enhanced by preparing a poorly relaxed system.

\section{Discussion and conclusion}
\label{sec:discussion}

To sum up, we have reproduced with numerical simulations the key experimental observations of Refs~\cite{Klongvessa2019_PRL,Klongvessa2019_PRE}.
In the ergodic supercooled regime, we confirm the mapping between the passive and active systems via the VTF relation of the density-dependence relaxation.
Remarkably, the mapping fails in the glass state and the structural relaxation of the system exhibits a nonmonotonic response to activity level that cannot be explained by the reentrant behavior of soft spheres.
Although we verified that the self-propulsion induces an instantaneous anisotropy of cage exploration, as predicted by DEAD model, we refute the predictions of the model regarding the influence of the rotational time.
By contrast, we demonstrate by varying the preparation protocol that the nonmonotonic response is linked to activity-enhanced aging.
This is consistent with our experimental observations that identified ergodicity breaking and the onset of the non monotonic behavior, both in glass~\cite{Klongvessa2019_PRL} and polycrystals~\cite{Klongvessa2019_PRE}.

From our results, we can conclude that low levels of self-propulsion are not significantly slowing down the way a Brownian particle explores its cage in steady state.
By contrast, low levels of self-propulsion during ageing are probably able to enhance the aging process, bringing faster the system to deeper minima of the energy landscape.
Thus an actively aged system has higher energy barriers to cross than its passively aged counterpart. 
When the propulsion force is not strong enough to compensate for the higher barrier, the active system is slower than the passive.
By contrast, when the propulsion force more than compensate for the higher barriers, the active system is faster.
Since the nonmonotonic behavior disappears at long persistence time, it seems that activity-enhanced aging needs relatively short persistence time, although this remains to be quantified.

We can relate this activated aging hypothesis with recent simulations showing nonmonotonic enhancement of steady-state relaxation in supercooled liquids~\cite{debets2021cage}. In particular, further simulation works should specifically look at how these enhanced steady-state relaxation translate or not into faster aging beyond ergodicity breaking.

%\input{sections/appendix}

% If in two-column mode, this environment will change to single-column format so that long equations can be displayed. 
%\begin{widetext}
%$$\mbox{put long equation here}$$
%\end{widetext}

% Figures should be put into the text as floats. 
% Use the graphics or graphicx packages (distributed with LaTeX2e).
% Put the label that you will use with \ref{} command in the braces of the \label{} command.
%
% \begin{figure}
% \end{figure}

% \begin{table}
% \end{tabular}
% \end{table}

\begin{acknowledgments}
% Put your acknowledgments here.
N.K. was supported by (i) Japan Society for the Promotion of Science (JSPS) (ii) Thailand Science Research and Innovation (TSRI), and (iii) National Science, Research and Innovation Fund (NSRF) of Thailand.
T.K. was supported by JSPS KAKENHI Grant Numbers JP20H00128, JP20H05157, JP19K03767, and JP18H01188.
We thank David Rodney for pointing to enhanced aging, and Olivier Dauchot for encouraging us in this direction.
N.K. and T.K. thank Kunimasa Miyazaki for fruitful discussions.

\end{acknowledgments}

\bibliography{mybib.bib}

\begin{thebibliography}{48}%
\makeatletter
\providecommand \@ifxundefined [1]{%
 \@ifx{#1\undefined}
}%
\providecommand \@ifnum [1]{%
 \ifnum #1\expandafter \@firstoftwo
 \else \expandafter \@secondoftwo
 \fi
}%
\providecommand \@ifx [1]{%
 \ifx #1\expandafter \@firstoftwo
 \else \expandafter \@secondoftwo
 \fi
}%
\providecommand \natexlab [1]{#1}%
\providecommand \enquote  [1]{``#1''}%
\providecommand \bibnamefont  [1]{#1}%
\providecommand \bibfnamefont [1]{#1}%
\providecommand \citenamefont [1]{#1}%
\providecommand \href@noop [0]{\@secondoftwo}%
\providecommand \href [0]{\begingroup \@sanitize@url \@href}%
\providecommand \@href[1]{\@@startlink{#1}\@@href}%
\providecommand \@@href[1]{\endgroup#1\@@endlink}%
\providecommand \@sanitize@url [0]{\catcode `\\12\catcode `\$12\catcode
  `\&12\catcode `\#12\catcode `\^12\catcode `\_12\catcode `\%12\relax}%
\providecommand \@@startlink[1]{}%
\providecommand \@@endlink[0]{}%
\providecommand \url  [0]{\begingroup\@sanitize@url \@url }%
\providecommand \@url [1]{\endgroup\@href {#1}{\urlprefix }}%
\providecommand \urlprefix  [0]{URL }%
\providecommand \Eprint [0]{\href }%
\providecommand \doibase [0]{http://dx.doi.org/}%
\providecommand \selectlanguage [0]{\@gobble}%
\providecommand \bibinfo  [0]{\@secondoftwo}%
\providecommand \bibfield  [0]{\@secondoftwo}%
\providecommand \translation [1]{[#1]}%
\providecommand \BibitemOpen [0]{}%
\providecommand \bibitemStop [0]{}%
\providecommand \bibitemNoStop [0]{.\EOS\space}%
\providecommand \EOS [0]{\spacefactor3000\relax}%
\providecommand \BibitemShut  [1]{\csname bibitem#1\endcsname}%
\let\auto@bib@innerbib\@empty
%</preamble>
\bibitem [{\citenamefont {Goldstein}(1969)}]{Goldstein1969}%
  \BibitemOpen
  \bibfield  {author} {\bibinfo {author} {\bibfnamefont {M.}~\bibnamefont
  {Goldstein}},\ }\bibfield  {title} {\enquote {\bibinfo {title} {Viscous
  {{Liquids}} and the {{Glass Transition}}: {{A Potential Energy Barrier
  Picture}}},}\ }\href {\doibase 10.1063/1.1672587} {\bibfield  {journal}
  {\bibinfo  {journal} {The Journal of Chemical Physics}\ }\textbf {\bibinfo
  {volume} {51}},\ \bibinfo {pages} {3728} (\bibinfo {year}
  {1969})}\BibitemShut {NoStop}%
\bibitem [{\citenamefont {Debenedetti}\ and\ \citenamefont
  {Stillinger}(2001)}]{Debenedetti2001}%
  \BibitemOpen
  \bibfield  {author} {\bibinfo {author} {\bibfnamefont {P.~G.}\ \bibnamefont
  {Debenedetti}}\ and\ \bibinfo {author} {\bibfnamefont {F.~H.}\ \bibnamefont
  {Stillinger}},\ }\bibfield  {title} {\enquote {\bibinfo {title} {Supercooled
  liquids and the glass transition},}\ }\href {\doibase 10.1038/35065704}
  {\bibfield  {journal} {\bibinfo  {journal} {Nature}\ }\textbf {\bibinfo
  {volume} {410}},\ \bibinfo {pages} {259--267} (\bibinfo {year}
  {2001})}\BibitemShut {NoStop}%
\bibitem [{\citenamefont {Berthier}\ and\ \citenamefont
  {Biroli}(2011{\natexlab{a}})}]{Berthier2011_review}%
  \BibitemOpen
  \bibfield  {author} {\bibinfo {author} {\bibfnamefont {L.}~\bibnamefont
  {Berthier}}\ and\ \bibinfo {author} {\bibfnamefont {G.}~\bibnamefont
  {Biroli}},\ }\bibfield  {title} {\enquote {\bibinfo {title} {Theoretical
  perspective on the glass transition and amorphous materials},}\ }\href
  {\doibase 10.1103/RevModPhys.83.587} {\bibfield  {journal} {\bibinfo
  {journal} {Rev. Mod. Phys.}\ }\textbf {\bibinfo {volume} {83}},\ \bibinfo
  {pages} {587--645} (\bibinfo {year} {2011}{\natexlab{a}})}\BibitemShut
  {NoStop}%
\bibitem [{\citenamefont {Charbonneau}\ \emph {et~al.}(2014)\citenamefont
  {Charbonneau}, \citenamefont {Kurchan}, \citenamefont {Parisi}, \citenamefont
  {Urbani},\ and\ \citenamefont {Zamponi}}]{Charbonneau2014}%
  \BibitemOpen
  \bibfield  {author} {\bibinfo {author} {\bibfnamefont {P.}~\bibnamefont
  {Charbonneau}}, \bibinfo {author} {\bibfnamefont {J.}~\bibnamefont
  {Kurchan}}, \bibinfo {author} {\bibfnamefont {G.}~\bibnamefont {Parisi}},
  \bibinfo {author} {\bibfnamefont {P.}~\bibnamefont {Urbani}}, \ and\ \bibinfo
  {author} {\bibfnamefont {F.}~\bibnamefont {Zamponi}},\ }\bibfield  {title}
  {\enquote {\bibinfo {title} {Fractal free energy landscapes in structural
  glasses},}\ }\href {\doibase 10.1038/ncomms4725} {\bibfield  {journal}
  {\bibinfo  {journal} {Nature Communications}\ }\textbf {\bibinfo {volume}
  {5}},\ \bibinfo {pages} {3725} (\bibinfo {year} {2014})}\BibitemShut
  {NoStop}%
\bibitem [{\citenamefont {Charbonneau}\ \emph {et~al.}(2017)\citenamefont
  {Charbonneau}, \citenamefont {Kurchan}, \citenamefont {Parisi}, \citenamefont
  {Urbani},\ and\ \citenamefont {Zamponi}}]{Charbonneau2017_review}%
  \BibitemOpen
  \bibfield  {author} {\bibinfo {author} {\bibfnamefont {P.}~\bibnamefont
  {Charbonneau}}, \bibinfo {author} {\bibfnamefont {J.}~\bibnamefont
  {Kurchan}}, \bibinfo {author} {\bibfnamefont {G.}~\bibnamefont {Parisi}},
  \bibinfo {author} {\bibfnamefont {P.}~\bibnamefont {Urbani}}, \ and\ \bibinfo
  {author} {\bibfnamefont {F.}~\bibnamefont {Zamponi}},\ }\bibfield  {title}
  {\enquote {\bibinfo {title} {Glass and jamming transitions: From exact
  results to finite-dimensional descriptions},}\ }\href {\doibase
  10.1146/annurev-conmatphys-031016-025334} {\bibfield  {journal} {\bibinfo
  {journal} {Annual Review of Condensed Matter Physics}\ }\textbf {\bibinfo
  {volume} {8}},\ \bibinfo {pages} {265--288} (\bibinfo {year} {2017})},\
  \Eprint
  {http://arxiv.org/abs/https://doi.org/10.1146/annurev-conmatphys-031016-025334}
  {https://doi.org/10.1146/annurev-conmatphys-031016-025334} \BibitemShut
  {NoStop}%
\bibitem [{\citenamefont {Struik}(1977)}]{struik1977physical}%
  \BibitemOpen
  \bibfield  {author} {\bibinfo {author} {\bibfnamefont {L.~C.~E.}\
  \bibnamefont {Struik}},\ }\bibfield  {title} {\enquote {\bibinfo {title}
  {Physical aging in amorphous polymers and other materials},}\ }\href@noop {}
  {\bibfield  {journal} {\bibinfo  {journal} {{PhD} Thesis Technische
  Hogeschool Delft}\ } (\bibinfo {year} {1977})}\BibitemShut {NoStop}%
\bibitem [{\citenamefont {Berthier}\ and\ \citenamefont
  {Biroli}(2011{\natexlab{b}})}]{berthier_theo_2011}%
  \BibitemOpen
  \bibfield  {author} {\bibinfo {author} {\bibfnamefont {L.}~\bibnamefont
  {Berthier}}\ and\ \bibinfo {author} {\bibfnamefont {G.}~\bibnamefont
  {Biroli}},\ }\bibfield  {title} {\enquote {\bibinfo {title} {Theoretical
  perspective on the glass transition and amorphous materials},}\ }\href
  {\doibase 10.1103/RevModPhys.83.587} {\bibfield  {journal} {\bibinfo
  {journal} {Rev. Mod. Phys.}\ }\textbf {\bibinfo {volume} {83}},\ \bibinfo
  {pages} {587--645} (\bibinfo {year} {2011}{\natexlab{b}})}\BibitemShut
  {NoStop}%
\bibitem [{\citenamefont {Vincent}\ \emph {et~al.}(1997)\citenamefont
  {Vincent}, \citenamefont {Hammann}, \citenamefont {Ocio}, \citenamefont
  {Bouchaud},\ and\ \citenamefont
  {Cugliandolo}}]{vincent1997SlowDynamicsAging}%
  \BibitemOpen
  \bibfield  {author} {\bibinfo {author} {\bibfnamefont {E.}~\bibnamefont
  {Vincent}}, \bibinfo {author} {\bibfnamefont {J.}~\bibnamefont {Hammann}},
  \bibinfo {author} {\bibfnamefont {M.}~\bibnamefont {Ocio}}, \bibinfo {author}
  {\bibfnamefont {J.-P.}\ \bibnamefont {Bouchaud}}, \ and\ \bibinfo {author}
  {\bibfnamefont {L.~F.}\ \bibnamefont {Cugliandolo}},\ }\bibfield  {title}
  {\enquote {\bibinfo {title} {Slow dynamics and aging in spin glasses},}\ }in\
  \href {\doibase 10.1007/BFb0104827} {\emph {\bibinfo {booktitle} {Complex
  {{Behaviour}} of {{Glassy Systems}}}}},\ \bibinfo {series and number}
  {Lecture {{Notes}} in {{Physics}}},\ \bibinfo {editor} {edited by\ \bibinfo
  {editor} {\bibfnamefont {M.}~\bibnamefont {Rub{\'i}}}\ and\ \bibinfo {editor}
  {\bibfnamefont {C.}~\bibnamefont {{P{\'e}rez-Vicente}}}}\ (\bibinfo
  {publisher} {{Springer}},\ \bibinfo {address} {{Berlin, Heidelberg}},\
  \bibinfo {year} {1997})\ pp.\ \bibinfo {pages} {184--219}\BibitemShut
  {NoStop}%
\bibitem [{\citenamefont {Scalliet}\ and\ \citenamefont
  {Berthier}(2019)}]{scalliet2019RejuvenationMemoryEffects}%
  \BibitemOpen
  \bibfield  {author} {\bibinfo {author} {\bibfnamefont {C.}~\bibnamefont
  {Scalliet}}\ and\ \bibinfo {author} {\bibfnamefont {L.}~\bibnamefont
  {Berthier}},\ }\bibfield  {title} {\enquote {\bibinfo {title} {Rejuvenation
  and {{Memory Effects}} in a {{Structural Glass}}},}\ }\href {\doibase
  10.1103/PhysRevLett.122.255502} {\bibfield  {journal} {\bibinfo  {journal}
  {Physical Review Letters}\ }\textbf {\bibinfo {volume} {122}},\ \bibinfo
  {pages} {255502} (\bibinfo {year} {2019})}\BibitemShut {NoStop}%
\bibitem [{\citenamefont {Viasnoff}\ and\ \citenamefont
  {Lequeux}(2002)}]{Viasnoff.2002}%
  \BibitemOpen
  \bibfield  {author} {\bibinfo {author} {\bibfnamefont {V.}~\bibnamefont
  {Viasnoff}}\ and\ \bibinfo {author} {\bibfnamefont {F.}~\bibnamefont
  {Lequeux}},\ }\bibfield  {title} {\enquote {\bibinfo {title} {{Rejuvenation
  and Overaging in a Colloidal Glass under Shear}},}\ }\href@noop {} {\bibfield
   {journal} {\bibinfo  {journal} {Physical Review Letters}\ }\textbf {\bibinfo
  {volume} {89}},\ \bibinfo {pages} {065701} (\bibinfo {year}
  {2002})}\BibitemShut {NoStop}%
\bibitem [{\citenamefont {Ikeda}, \citenamefont {Berthier},\ and\ \citenamefont
  {Sollich}(2013)}]{Ikeda2013a}%
  \BibitemOpen
  \bibfield  {author} {\bibinfo {author} {\bibfnamefont {A.}~\bibnamefont
  {Ikeda}}, \bibinfo {author} {\bibfnamefont {L.}~\bibnamefont {Berthier}}, \
  and\ \bibinfo {author} {\bibfnamefont {P.}~\bibnamefont {Sollich}},\
  }\bibfield  {title} {\enquote {\bibinfo {title} {Disentangling glass and
  jamming physics in the rheology of soft materials},}\ }\href {\doibase
  10.1039/c3sm50503k} {\bibfield  {journal} {\bibinfo  {journal} {Soft Matter}\
  }\textbf {\bibinfo {volume} {9}},\ \bibinfo {pages} {7669} (\bibinfo {year}
  {2013})}\BibitemShut {NoStop}%
\bibitem [{\citenamefont {Janssen}, \citenamefont {Kaiser},\ and\ \citenamefont
  {L{\"o}wen}(2017)}]{janssen2017AgingRejuvenationActive}%
  \BibitemOpen
  \bibfield  {author} {\bibinfo {author} {\bibfnamefont {L.~M.~C.}\
  \bibnamefont {Janssen}}, \bibinfo {author} {\bibfnamefont {A.}~\bibnamefont
  {Kaiser}}, \ and\ \bibinfo {author} {\bibfnamefont {H.}~\bibnamefont
  {L{\"o}wen}},\ }\bibfield  {title} {\enquote {\bibinfo {title} {Aging and
  rejuvenation of active matter under topological constraints},}\ }\href
  {\doibase 10.1038/s41598-017-05569-6} {\bibfield  {journal} {\bibinfo
  {journal} {Scientific Reports}\ }\textbf {\bibinfo {volume} {7}},\ \bibinfo
  {pages} {5667} (\bibinfo {year} {2017})}\BibitemShut {NoStop}%
\bibitem [{\citenamefont {Janssen}(2019)}]{Janssen2019_ActiveGlass}%
  \BibitemOpen
  \bibfield  {author} {\bibinfo {author} {\bibfnamefont {L.~M.~C.}\
  \bibnamefont {Janssen}},\ }\bibfield  {title} {\enquote {\bibinfo {title}
  {Active glasses},}\ }\href {\doibase 10.1088/1361-648x/ab3e90} {\bibfield
  {journal} {\bibinfo  {journal} {Journal of Physics: Condensed Matter}\
  }\textbf {\bibinfo {volume} {31}},\ \bibinfo {pages} {503002} (\bibinfo
  {year} {2019})}\BibitemShut {NoStop}%
\bibitem [{\citenamefont {Garcia}\ \emph {et~al.}(2015)\citenamefont {Garcia},
  \citenamefont {Hannezo}, \citenamefont {Elgeti}, \citenamefont {Joanny},
  \citenamefont {Silberzan},\ and\ \citenamefont
  {Gov}}]{Garcia2015_cell_activejamming}%
  \BibitemOpen
  \bibfield  {author} {\bibinfo {author} {\bibfnamefont {S.}~\bibnamefont
  {Garcia}}, \bibinfo {author} {\bibfnamefont {E.}~\bibnamefont {Hannezo}},
  \bibinfo {author} {\bibfnamefont {J.}~\bibnamefont {Elgeti}}, \bibinfo
  {author} {\bibfnamefont {J.-F.}\ \bibnamefont {Joanny}}, \bibinfo {author}
  {\bibfnamefont {P.}~\bibnamefont {Silberzan}}, \ and\ \bibinfo {author}
  {\bibfnamefont {N.~S.}\ \bibnamefont {Gov}},\ }\bibfield  {title} {\enquote
  {\bibinfo {title} {Physics of active jamming during collective cellular
  motion in a monolayer},}\ }\href {\doibase 10.1073/pnas.1510973112}
  {\bibfield  {journal} {\bibinfo  {journal} {Proceedings of the National
  Academy of Sciences}\ }\textbf {\bibinfo {volume} {112}},\ \bibinfo {pages}
  {15314--15319} (\bibinfo {year} {2015})},\ \Eprint
  {http://arxiv.org/abs/https://www.pnas.org/content/112/50/15314.full.pdf}
  {https://www.pnas.org/content/112/50/15314.full.pdf} \BibitemShut {NoStop}%
\bibitem [{\citenamefont {Bi}\ \emph {et~al.}(2016)\citenamefont {Bi},
  \citenamefont {Yang}, \citenamefont {Marchetti},\ and\ \citenamefont
  {Manning}}]{Bi2016PRX}%
  \BibitemOpen
  \bibfield  {author} {\bibinfo {author} {\bibfnamefont {D.}~\bibnamefont
  {Bi}}, \bibinfo {author} {\bibfnamefont {X.}~\bibnamefont {Yang}}, \bibinfo
  {author} {\bibfnamefont {M.~C.}\ \bibnamefont {Marchetti}}, \ and\ \bibinfo
  {author} {\bibfnamefont {M.~L.}\ \bibnamefont {Manning}},\ }\bibfield
  {title} {\enquote {\bibinfo {title} {Motility-driven glass and jamming
  transitions in biological tissues},}\ }\href {\doibase
  10.1103/PhysRevX.6.021011} {\bibfield  {journal} {\bibinfo  {journal} {Phys.
  Rev. X}\ }\textbf {\bibinfo {volume} {6}},\ \bibinfo {pages} {021011}
  (\bibinfo {year} {2016})}\BibitemShut {NoStop}%
\bibitem [{\citenamefont {Ramaswamy}(2010)}]{Ramaswamy2010}%
  \BibitemOpen
  \bibfield  {author} {\bibinfo {author} {\bibfnamefont {S.}~\bibnamefont
  {Ramaswamy}},\ }\bibfield  {title} {\enquote {\bibinfo {title} {The mechanics
  and statistics of active matter},}\ }\href {\doibase
  10.1146/annurev-conmatphys-070909-104101} {\bibfield  {journal} {\bibinfo
  {journal} {Annual Review of Condensed Matter Physics}\ }\textbf {\bibinfo
  {volume} {1}},\ \bibinfo {pages} {323--345} (\bibinfo {year} {2010})},\
  \Eprint
  {http://arxiv.org/abs/https://doi.org/10.1146/annurev-conmatphys-070909-104101}
  {https://doi.org/10.1146/annurev-conmatphys-070909-104101} \BibitemShut
  {NoStop}%
\bibitem [{\citenamefont {Vicsek}\ and\ \citenamefont
  {Zafeiris}(2012)}]{Vicsek_review2012}%
  \BibitemOpen
  \bibfield  {author} {\bibinfo {author} {\bibfnamefont {T.}~\bibnamefont
  {Vicsek}}\ and\ \bibinfo {author} {\bibfnamefont {A.}~\bibnamefont
  {Zafeiris}},\ }\bibfield  {title} {\enquote {\bibinfo {title} {Collective
  motion},}\ }\href {\doibase 10.1016/j.physrep.2012.03.004} {\bibfield
  {journal} {\bibinfo  {journal} {Physics Reports}\ }\textbf {\bibinfo {volume}
  {517}},\ \bibinfo {pages} {71--140} (\bibinfo {year} {2012})}\BibitemShut
  {NoStop}%
\bibitem [{\citenamefont {Marchetti}\ \emph {et~al.}(2013)\citenamefont
  {Marchetti}, \citenamefont {Joanny}, \citenamefont {Ramaswamy}, \citenamefont
  {Liverpool}, \citenamefont {Prost}, \citenamefont {Rao},\ and\ \citenamefont
  {Simha}}]{Marchetti2013}%
  \BibitemOpen
  \bibfield  {author} {\bibinfo {author} {\bibfnamefont {M.~C.}\ \bibnamefont
  {Marchetti}}, \bibinfo {author} {\bibfnamefont {J.~F.}\ \bibnamefont
  {Joanny}}, \bibinfo {author} {\bibfnamefont {S.}~\bibnamefont {Ramaswamy}},
  \bibinfo {author} {\bibfnamefont {T.~B.}\ \bibnamefont {Liverpool}}, \bibinfo
  {author} {\bibfnamefont {J.}~\bibnamefont {Prost}}, \bibinfo {author}
  {\bibfnamefont {M.}~\bibnamefont {Rao}}, \ and\ \bibinfo {author}
  {\bibfnamefont {R.~A.}\ \bibnamefont {Simha}},\ }\bibfield  {title} {\enquote
  {\bibinfo {title} {Hydrodynamics of soft active matter},}\ }\href {\doibase
  10.1103/revmodphys.85.1143} {\bibfield  {journal} {\bibinfo  {journal}
  {Reviews of Modern Physics}\ }\textbf {\bibinfo {volume} {85}},\ \bibinfo
  {pages} {1143--1189} (\bibinfo {year} {2013})}\BibitemShut {NoStop}%
\bibitem [{\citenamefont {Bechinger}\ \emph {et~al.}(2016)\citenamefont
  {Bechinger}, \citenamefont {Di~Leonardo}, \citenamefont {L\"owen},
  \citenamefont {Reichhardt}, \citenamefont {Volpe},\ and\ \citenamefont
  {Volpe}}]{Bechinger2016}%
  \BibitemOpen
  \bibfield  {author} {\bibinfo {author} {\bibfnamefont {C.}~\bibnamefont
  {Bechinger}}, \bibinfo {author} {\bibfnamefont {R.}~\bibnamefont
  {Di~Leonardo}}, \bibinfo {author} {\bibfnamefont {H.}~\bibnamefont
  {L\"owen}}, \bibinfo {author} {\bibfnamefont {C.}~\bibnamefont {Reichhardt}},
  \bibinfo {author} {\bibfnamefont {G.}~\bibnamefont {Volpe}}, \ and\ \bibinfo
  {author} {\bibfnamefont {G.}~\bibnamefont {Volpe}},\ }\bibfield  {title}
  {\enquote {\bibinfo {title} {Active particles in complex and crowded
  environments},}\ }\href {\doibase 10.1103/RevModPhys.88.045006} {\bibfield
  {journal} {\bibinfo  {journal} {Rev. Mod. Phys.}\ }\textbf {\bibinfo {volume}
  {88}},\ \bibinfo {pages} {045006} (\bibinfo {year} {2016})}\BibitemShut
  {NoStop}%
\bibitem [{\citenamefont {Howse}\ \emph {et~al.}(2007)\citenamefont {Howse},
  \citenamefont {Jones}, \citenamefont {Ryan}, \citenamefont {Gough},
  \citenamefont {Vafabakhsh},\ and\ \citenamefont
  {Golestanian}}]{Howse_PRL_2007}%
  \BibitemOpen
  \bibfield  {author} {\bibinfo {author} {\bibfnamefont {J.~R.}\ \bibnamefont
  {Howse}}, \bibinfo {author} {\bibfnamefont {R.~A.~L.}\ \bibnamefont {Jones}},
  \bibinfo {author} {\bibfnamefont {A.~J.}\ \bibnamefont {Ryan}}, \bibinfo
  {author} {\bibfnamefont {T.}~\bibnamefont {Gough}}, \bibinfo {author}
  {\bibfnamefont {R.}~\bibnamefont {Vafabakhsh}}, \ and\ \bibinfo {author}
  {\bibfnamefont {R.}~\bibnamefont {Golestanian}},\ }\bibfield  {title}
  {\enquote {\bibinfo {title} {Self-motile colloidal particles: From directed
  propulsion to random walk},}\ }\href {\doibase 10.1103/PhysRevLett.99.048102}
  {\bibfield  {journal} {\bibinfo  {journal} {Phys. Rev. Lett.}\ }\textbf
  {\bibinfo {volume} {99}},\ \bibinfo {pages} {048102} (\bibinfo {year}
  {2007})}\BibitemShut {NoStop}%
\bibitem [{\citenamefont {Tailleur}\ and\ \citenamefont
  {Cates}(2009)}]{Tailleur2009}%
  \BibitemOpen
  \bibfield  {author} {\bibinfo {author} {\bibfnamefont {J.}~\bibnamefont
  {Tailleur}}\ and\ \bibinfo {author} {\bibfnamefont {M.~E.}\ \bibnamefont
  {Cates}},\ }\bibfield  {title} {\enquote {\bibinfo {title} {Sedimentation,
  trapping, and rectification of dilute bacteria},}\ }\href
  {http://stacks.iop.org/0295-5075/86/i=6/a=60002} {\bibfield  {journal}
  {\bibinfo  {journal} {EPL}\ }\textbf {\bibinfo {volume} {86}},\ \bibinfo
  {pages} {60002} (\bibinfo {year} {2009})}\BibitemShut {NoStop}%
\bibitem [{\citenamefont {Palacci}\ \emph {et~al.}(2010)\citenamefont
  {Palacci}, \citenamefont {Cottin-Bizonne}, \citenamefont {Ybert},\ and\
  \citenamefont {Bocquet}}]{Palacci2010}%
  \BibitemOpen
  \bibfield  {author} {\bibinfo {author} {\bibfnamefont {J.}~\bibnamefont
  {Palacci}}, \bibinfo {author} {\bibfnamefont {C.}~\bibnamefont
  {Cottin-Bizonne}}, \bibinfo {author} {\bibfnamefont {C.}~\bibnamefont
  {Ybert}}, \ and\ \bibinfo {author} {\bibfnamefont {L.}~\bibnamefont
  {Bocquet}},\ }\bibfield  {title} {\enquote {\bibinfo {title} {Sedimentation
  and effective temperature of active colloidal suspensions},}\ }\href
  {\doibase 10.1103/PhysRevLett.105.088304} {\bibfield  {journal} {\bibinfo
  {journal} {Phys. Rev. Lett.}\ }\textbf {\bibinfo {volume} {105}},\ \bibinfo
  {pages} {088304} (\bibinfo {year} {2010})}\BibitemShut {NoStop}%
\bibitem [{\citenamefont {Ginot}\ \emph {et~al.}(2015)\citenamefont {Ginot},
  \citenamefont {Theurkauff}, \citenamefont {Levis}, \citenamefont {Ybert},
  \citenamefont {Bocquet}, \citenamefont {Berthier},\ and\ \citenamefont
  {Cottin-Bizonne}}]{Ginot2015}%
  \BibitemOpen
  \bibfield  {author} {\bibinfo {author} {\bibfnamefont {F.}~\bibnamefont
  {Ginot}}, \bibinfo {author} {\bibfnamefont {I.}~\bibnamefont {Theurkauff}},
  \bibinfo {author} {\bibfnamefont {D.}~\bibnamefont {Levis}}, \bibinfo
  {author} {\bibfnamefont {C.}~\bibnamefont {Ybert}}, \bibinfo {author}
  {\bibfnamefont {L.}~\bibnamefont {Bocquet}}, \bibinfo {author} {\bibfnamefont
  {L.}~\bibnamefont {Berthier}}, \ and\ \bibinfo {author} {\bibfnamefont
  {C.}~\bibnamefont {Cottin-Bizonne}},\ }\bibfield  {title} {\enquote {\bibinfo
  {title} {Nonequilibrium equation of state in suspensions of active
  colloids},}\ }\href {\doibase 10.1103/PhysRevX.5.011004} {\bibfield
  {journal} {\bibinfo  {journal} {Phys. Rev. X}\ }\textbf {\bibinfo {volume}
  {5}},\ \bibinfo {pages} {011004} (\bibinfo {year} {2015})}\BibitemShut
  {NoStop}%
\bibitem [{\citenamefont {Klongvessa}\ \emph
  {et~al.}(2019{\natexlab{a}})\citenamefont {Klongvessa}, \citenamefont
  {Ginot}, \citenamefont {Ybert}, \citenamefont {Cottin-Bizonne},\ and\
  \citenamefont {Leocmach}}]{Klongvessa2019_PRL}%
  \BibitemOpen
  \bibfield  {author} {\bibinfo {author} {\bibfnamefont {N.}~\bibnamefont
  {Klongvessa}}, \bibinfo {author} {\bibfnamefont {F.}~\bibnamefont {Ginot}},
  \bibinfo {author} {\bibfnamefont {C.}~\bibnamefont {Ybert}}, \bibinfo
  {author} {\bibfnamefont {C.}~\bibnamefont {Cottin-Bizonne}}, \ and\ \bibinfo
  {author} {\bibfnamefont {M.}~\bibnamefont {Leocmach}},\ }\bibfield  {title}
  {\enquote {\bibinfo {title} {Active glass: Ergodicity breaking dramatically
  affects response to self-propulsion},}\ }\href {\doibase
  10.1103/PhysRevLett.123.248004} {\bibfield  {journal} {\bibinfo  {journal}
  {Phys. Rev. Lett.}\ }\textbf {\bibinfo {volume} {123}},\ \bibinfo {pages}
  {248004} (\bibinfo {year} {2019}{\natexlab{a}})}\BibitemShut {NoStop}%
\bibitem [{\citenamefont {Klongvessa}\ \emph
  {et~al.}(2019{\natexlab{b}})\citenamefont {Klongvessa}, \citenamefont
  {Ginot}, \citenamefont {Ybert}, \citenamefont {Cottin-Bizonne},\ and\
  \citenamefont {Leocmach}}]{Klongvessa2019_PRE}%
  \BibitemOpen
  \bibfield  {author} {\bibinfo {author} {\bibfnamefont {N.}~\bibnamefont
  {Klongvessa}}, \bibinfo {author} {\bibfnamefont {F.}~\bibnamefont {Ginot}},
  \bibinfo {author} {\bibfnamefont {C.}~\bibnamefont {Ybert}}, \bibinfo
  {author} {\bibfnamefont {C.}~\bibnamefont {Cottin-Bizonne}}, \ and\ \bibinfo
  {author} {\bibfnamefont {M.}~\bibnamefont {Leocmach}},\ }\bibfield  {title}
  {\enquote {\bibinfo {title} {Nonmonotonic behavior in dense assemblies of
  active colloids},}\ }\href {\doibase 10.1103/PhysRevE.100.062603} {\bibfield
  {journal} {\bibinfo  {journal} {Phys. Rev. E}\ }\textbf {\bibinfo {volume}
  {100}},\ \bibinfo {pages} {062603} (\bibinfo {year}
  {2019}{\natexlab{b}})}\BibitemShut {NoStop}%
\bibitem [{\citenamefont {Liluashvili}, \citenamefont {{\'O}nody},\ and\
  \citenamefont {Voigtmann}(2017)}]{liluashviliModecouplingTheoryActive2017}%
  \BibitemOpen
  \bibfield  {author} {\bibinfo {author} {\bibfnamefont {A.}~\bibnamefont
  {Liluashvili}}, \bibinfo {author} {\bibfnamefont {J.}~\bibnamefont
  {{\'O}nody}}, \ and\ \bibinfo {author} {\bibfnamefont {T.}~\bibnamefont
  {Voigtmann}},\ }\bibfield  {title} {\enquote {\bibinfo {title} {Mode-coupling
  theory for active {{Brownian}} particles},}\ }\href {\doibase
  10.1103/PhysRevE.96.062608} {\bibfield  {journal} {\bibinfo  {journal}
  {Physical Review E}\ }\textbf {\bibinfo {volume} {96}},\ \bibinfo {pages}
  {062608} (\bibinfo {year} {2017})}\BibitemShut {NoStop}%
\bibitem [{\citenamefont {Debets}, \citenamefont {{de Wit}},\ and\
  \citenamefont {Janssen}(2021)}]{debets2021cage}%
  \BibitemOpen
  \bibfield  {author} {\bibinfo {author} {\bibfnamefont {V.~E.}\ \bibnamefont
  {Debets}}, \bibinfo {author} {\bibfnamefont {X.~M.}\ \bibnamefont {{de
  Wit}}}, \ and\ \bibinfo {author} {\bibfnamefont {L.~M.~C.}\ \bibnamefont
  {Janssen}},\ }\bibfield  {title} {\enquote {\bibinfo {title} {Cage {{Length
  Controls}} the {{Nonmonotonic Dynamics}} of {{Active Glassy Matter}}},}\
  }\href {\doibase 10.1103/PhysRevLett.127.278002} {\bibfield  {journal}
  {\bibinfo  {journal} {Physical Review Letters}\ }\textbf {\bibinfo {volume}
  {127}},\ \bibinfo {pages} {278002} (\bibinfo {year} {2021})},\ \Eprint
  {http://arxiv.org/abs/2111.11171} {arXiv:2111.11171} \BibitemShut {NoStop}%
\bibitem [{\citenamefont {Berthier}, \citenamefont {Flenner},\ and\
  \citenamefont {Szamel}(2017)}]{Berthier2017}%
  \BibitemOpen
  \bibfield  {author} {\bibinfo {author} {\bibfnamefont {L.}~\bibnamefont
  {Berthier}}, \bibinfo {author} {\bibfnamefont {E.}~\bibnamefont {Flenner}}, \
  and\ \bibinfo {author} {\bibfnamefont {G.}~\bibnamefont {Szamel}},\
  }\bibfield  {title} {\enquote {\bibinfo {title} {How active forces influence
  nonequilibrium glass transitions},}\ }\href
  {http://stacks.iop.org/1367-2630/19/i=12/a=125006} {\bibfield  {journal}
  {\bibinfo  {journal} {New J. Phys.}\ }\textbf {\bibinfo {volume} {19}},\
  \bibinfo {pages} {125006} (\bibinfo {year} {2017})}\BibitemShut {NoStop}%
\bibitem [{\citenamefont {Nandi}\ \emph {et~al.}(2018)\citenamefont {Nandi},
  \citenamefont {Mandal}, \citenamefont {Bhuyan}, \citenamefont {Dasgupta},
  \citenamefont {Rao},\ and\ \citenamefont {Gov}}]{Nandi2018}%
  \BibitemOpen
  \bibfield  {author} {\bibinfo {author} {\bibfnamefont {S.~K.}\ \bibnamefont
  {Nandi}}, \bibinfo {author} {\bibfnamefont {R.}~\bibnamefont {Mandal}},
  \bibinfo {author} {\bibfnamefont {P.~J.}\ \bibnamefont {Bhuyan}}, \bibinfo
  {author} {\bibfnamefont {C.}~\bibnamefont {Dasgupta}}, \bibinfo {author}
  {\bibfnamefont {M.}~\bibnamefont {Rao}}, \ and\ \bibinfo {author}
  {\bibfnamefont {N.~S.}\ \bibnamefont {Gov}},\ }\bibfield  {title} {\enquote
  {\bibinfo {title} {A random first-order transition theory for an active
  glass},}\ }\href {\doibase 10.1073/pnas.1721324115} {\bibfield  {journal}
  {\bibinfo  {journal} {Proceedings of the National Academy of Sciences}\
  }\textbf {\bibinfo {volume} {115}},\ \bibinfo {pages} {7688--7693} (\bibinfo
  {year} {2018})},\ \Eprint
  {http://arxiv.org/abs/https://www.pnas.org/content/115/30/7688.full.pdf}
  {https://www.pnas.org/content/115/30/7688.full.pdf} \BibitemShut {NoStop}%
\bibitem [{\citenamefont {Mandal}\ \emph {et~al.}(2020)\citenamefont {Mandal},
  \citenamefont {Bhuyan}, \citenamefont {Chaudhuri}, \citenamefont {Dasgupta},\
  and\ \citenamefont {Rao}}]{mandal_extreme_2020}%
  \BibitemOpen
  \bibfield  {author} {\bibinfo {author} {\bibfnamefont {R.}~\bibnamefont
  {Mandal}}, \bibinfo {author} {\bibfnamefont {P.~J.}\ \bibnamefont {Bhuyan}},
  \bibinfo {author} {\bibfnamefont {P.}~\bibnamefont {Chaudhuri}}, \bibinfo
  {author} {\bibfnamefont {C.}~\bibnamefont {Dasgupta}}, \ and\ \bibinfo
  {author} {\bibfnamefont {M.}~\bibnamefont {Rao}},\ }\bibfield  {title}
  {\enquote {\bibinfo {title} {Extreme active matter at high densities},}\
  }\href {\doibase 10.1038/s41467-020-16130-x} {\bibfield  {journal} {\bibinfo
  {journal} {Nature Communications}\ }\textbf {\bibinfo {volume} {11}},\
  \bibinfo {pages} {2581} (\bibinfo {year} {2020})}\BibitemShut {NoStop}%
\bibitem [{\citenamefont {Mandal}\ and\ \citenamefont
  {Sollich}(2020)}]{Mandal_Sollich_PRL2020_multi_aging}%
  \BibitemOpen
  \bibfield  {author} {\bibinfo {author} {\bibfnamefont {R.}~\bibnamefont
  {Mandal}}\ and\ \bibinfo {author} {\bibfnamefont {P.}~\bibnamefont
  {Sollich}},\ }\bibfield  {title} {\enquote {\bibinfo {title} {Multiple types
  of aging in active glasses},}\ }\href {\doibase
  10.1103/PhysRevLett.125.218001} {\bibfield  {journal} {\bibinfo  {journal}
  {Phys. Rev. Lett.}\ }\textbf {\bibinfo {volume} {125}},\ \bibinfo {pages}
  {218001} (\bibinfo {year} {2020})}\BibitemShut {NoStop}%
\bibitem [{\citenamefont {Janzen}\ and\ \citenamefont
  {Janssen}(2021)}]{janzen2021agingthermalglass}%
  \BibitemOpen
  \bibfield  {author} {\bibinfo {author} {\bibfnamefont {G.}~\bibnamefont
  {Janzen}}\ and\ \bibinfo {author} {\bibfnamefont {L.~M.~C.}\ \bibnamefont
  {Janssen}},\ }\href@noop {} {\enquote {\bibinfo {title} {Aging in thermal
  active glasses},}\ } (\bibinfo {year} {2021}),\ \Eprint
  {http://arxiv.org/abs/2105.05705} {arXiv:2105.05705 [cond-mat.stat-mech]}
  \BibitemShut {NoStop}%
\bibitem [{\citenamefont {Paul}, \citenamefont {Nandi},\ and\ \citenamefont
  {Karmakar}(2021)}]{paul_DH_activeglass_2021}%
  \BibitemOpen
  \bibfield  {author} {\bibinfo {author} {\bibfnamefont {K.}~\bibnamefont
  {Paul}}, \bibinfo {author} {\bibfnamefont {S.~K.}\ \bibnamefont {Nandi}}, \
  and\ \bibinfo {author} {\bibfnamefont {S.}~\bibnamefont {Karmakar}},\
  }\bibfield  {title} {\enquote {\bibinfo {title} {Dynamic heterogeneity in
  active glass-forming liquids is qualitatively different compared to its
  equilibrium behaviour},}\ }\href {http://arxiv.org/abs/2105.12702} {\bibfield
   {journal} {\bibinfo  {journal} {arXiv:2105.12702 [cond-mat]}\ } (\bibinfo
  {year} {2021})},\ \bibinfo {note} {arXiv: 2105.12702}\BibitemShut {NoStop}%
\bibitem [{\citenamefont {Shiba}\ \emph {et~al.}(2016)\citenamefont {Shiba},
  \citenamefont {Yamada}, \citenamefont {Kawasaki},\ and\ \citenamefont
  {Kim}}]{shiba2016UnveilingDimensionalityDependence}%
  \BibitemOpen
  \bibfield  {author} {\bibinfo {author} {\bibfnamefont {H.}~\bibnamefont
  {Shiba}}, \bibinfo {author} {\bibfnamefont {Y.}~\bibnamefont {Yamada}},
  \bibinfo {author} {\bibfnamefont {T.}~\bibnamefont {Kawasaki}}, \ and\
  \bibinfo {author} {\bibfnamefont {K.}~\bibnamefont {Kim}},\ }\bibfield
  {title} {\enquote {\bibinfo {title} {Unveiling {{Dimensionality Dependence}}
  of {{Glassy Dynamics}}: {{2D Infinite Fluctuation Eclipses Inherent
  Structural Relaxation}}},}\ }\href {\doibase 10.1103/PhysRevLett.117.245701}
  {\bibfield  {journal} {\bibinfo  {journal} {Physical Review Letters}\
  }\textbf {\bibinfo {volume} {117}},\ \bibinfo {pages} {245701} (\bibinfo
  {year} {2016})}\BibitemShut {NoStop}%
\bibitem [{\citenamefont {Illing}\ \emph {et~al.}(2017)\citenamefont {Illing},
  \citenamefont {Fritschi}, \citenamefont {Kaiser}, \citenamefont {Klix},
  \citenamefont {Maret},\ and\ \citenamefont
  {Keim}}]{illing2017MerminWagnerFluctuations}%
  \BibitemOpen
  \bibfield  {author} {\bibinfo {author} {\bibfnamefont {B.}~\bibnamefont
  {Illing}}, \bibinfo {author} {\bibfnamefont {S.}~\bibnamefont {Fritschi}},
  \bibinfo {author} {\bibfnamefont {H.}~\bibnamefont {Kaiser}}, \bibinfo
  {author} {\bibfnamefont {C.~L.}\ \bibnamefont {Klix}}, \bibinfo {author}
  {\bibfnamefont {G.}~\bibnamefont {Maret}}, \ and\ \bibinfo {author}
  {\bibfnamefont {P.}~\bibnamefont {Keim}},\ }\bibfield  {title} {\enquote
  {\bibinfo {title} {Mermin\textendash{{Wagner}} fluctuations in {{2D}}
  amorphous solids},}\ }\href {\doibase 10.1073/pnas.1612964114} {\bibfield
  {journal} {\bibinfo  {journal} {Proceedings of the National Academy of
  Sciences}\ }\textbf {\bibinfo {volume} {114}},\ \bibinfo {pages} {1856--1861}
  (\bibinfo {year} {2017})}\BibitemShut {NoStop}%
\bibitem [{\citenamefont {Fily}\ and\ \citenamefont
  {Marchetti}(2012)}]{Fily2012}%
  \BibitemOpen
  \bibfield  {author} {\bibinfo {author} {\bibfnamefont {Y.}~\bibnamefont
  {Fily}}\ and\ \bibinfo {author} {\bibfnamefont {M.~C.}\ \bibnamefont
  {Marchetti}},\ }\bibfield  {title} {\enquote {\bibinfo {title} {Athermal
  phase separation of self-propelled particles with no alignment},}\ }\href
  {\doibase 10.1103/physrevlett.108.235702} {\bibfield  {journal} {\bibinfo
  {journal} {Phys. Rev. Lett.}\ }\textbf {\bibinfo {volume} {108}} (\bibinfo
  {year} {2012}),\ 10.1103/physrevlett.108.235702}\BibitemShut {NoStop}%
\bibitem [{\citenamefont {Volpe}, \citenamefont {Gigan},\ and\ \citenamefont
  {Volpe}(2014)}]{Volpe2014}%
  \BibitemOpen
  \bibfield  {author} {\bibinfo {author} {\bibfnamefont {G.}~\bibnamefont
  {Volpe}}, \bibinfo {author} {\bibfnamefont {S.}~\bibnamefont {Gigan}}, \ and\
  \bibinfo {author} {\bibfnamefont {G.}~\bibnamefont {Volpe}},\ }\bibfield
  {title} {\enquote {\bibinfo {title} {Simulation of the active brownian motion
  of a microswimmer},}\ }\href {\doibase 10.1119/1.4870398} {\bibfield
  {journal} {\bibinfo  {journal} {American Journal of Physics}\ }\textbf
  {\bibinfo {volume} {82}},\ \bibinfo {pages} {659--664} (\bibinfo {year}
  {2014})},\ \Eprint {http://arxiv.org/abs/https://doi.org/10.1119/1.4870398}
  {https://doi.org/10.1119/1.4870398} \BibitemShut {NoStop}%
\bibitem [{\citenamefont {O'Hern}\ \emph {et~al.}(2002)\citenamefont {O'Hern},
  \citenamefont {Langer}, \citenamefont {Liu},\ and\ \citenamefont
  {Nagel}}]{OHern2002}%
  \BibitemOpen
  \bibfield  {author} {\bibinfo {author} {\bibfnamefont {C.~S.}\ \bibnamefont
  {O'Hern}}, \bibinfo {author} {\bibfnamefont {S.}~\bibnamefont {Langer}},
  \bibinfo {author} {\bibfnamefont {A.~J.}\ \bibnamefont {Liu}}, \ and\
  \bibinfo {author} {\bibfnamefont {S.}~\bibnamefont {Nagel}},\ }\bibfield
  {title} {\enquote {\bibinfo {title} {Random {{Packings}} of {{Frictionless
  Particles}}},}\ }\href {\doibase 10.1103/PhysRevLett.88.075507} {\bibfield
  {journal} {\bibinfo  {journal} {Physical Review Letters}\ }\textbf {\bibinfo
  {volume} {88}},\ \bibinfo {pages} {075507} (\bibinfo {year}
  {2002})}\BibitemShut {NoStop}%
\bibitem [{\citenamefont {Berthier}\ and\ \citenamefont
  {Witten}(2009)}]{Berthier2009}%
  \BibitemOpen
  \bibfield  {author} {\bibinfo {author} {\bibfnamefont {L.}~\bibnamefont
  {Berthier}}\ and\ \bibinfo {author} {\bibfnamefont {T.}~\bibnamefont
  {Witten}},\ }\bibfield  {title} {\enquote {\bibinfo {title} {Compressing
  nearly hard sphere fluids increases glass fragility},}\ }\href {\doibase
  10.1209/0295-5075/86/10001} {\bibfield  {journal} {\bibinfo  {journal} {EPL
  (Europhysics Letters)}\ }\textbf {\bibinfo {volume} {86}},\ \bibinfo {pages}
  {10001} (\bibinfo {year} {2009})}\BibitemShut {NoStop}%
\bibitem [{\citenamefont {Kawasaki}\ and\ \citenamefont
  {Tanaka}(2014)}]{kawasaki2014StructuralEvolutionAging}%
  \BibitemOpen
  \bibfield  {author} {\bibinfo {author} {\bibfnamefont {T.}~\bibnamefont
  {Kawasaki}}\ and\ \bibinfo {author} {\bibfnamefont {H.}~\bibnamefont
  {Tanaka}},\ }\bibfield  {title} {\enquote {\bibinfo {title} {Structural
  evolution in the aging process of supercooled colloidal liquids},}\ }\href
  {\doibase 10.1103/PhysRevE.89.062315} {\bibfield  {journal} {\bibinfo
  {journal} {Physical Review E}\ }\textbf {\bibinfo {volume} {89}},\ \bibinfo
  {pages} {062315} (\bibinfo {year} {2014})}\BibitemShut {NoStop}%
\bibitem [{\citenamefont {Gomez-Solano}\ \emph {et~al.}(2017)\citenamefont
  {Gomez-Solano}, \citenamefont {Samin}, \citenamefont {Lozano}, \citenamefont
  {Ruedas-Batuecas}, \citenamefont {Roij},\ and\ \citenamefont
  {Bechinger}}]{Gomez-Solano.2017}%
  \BibitemOpen
  \bibfield  {author} {\bibinfo {author} {\bibfnamefont {J.~R.}\ \bibnamefont
  {Gomez-Solano}}, \bibinfo {author} {\bibfnamefont {S.}~\bibnamefont {Samin}},
  \bibinfo {author} {\bibfnamefont {C.}~\bibnamefont {Lozano}}, \bibinfo
  {author} {\bibfnamefont {P.}~\bibnamefont {Ruedas-Batuecas}}, \bibinfo
  {author} {\bibfnamefont {R.~V.}\ \bibnamefont {Roij}}, \ and\ \bibinfo
  {author} {\bibfnamefont {C.}~\bibnamefont {Bechinger}},\ }\bibfield  {title}
  {\enquote {\bibinfo {title} {{Tuning the motility and directionality of
  self-propelled colloids}},}\ }\href {\doibase 10.1038/s41598-017-14126-0}
  {\bibfield  {journal} {\bibinfo  {journal} {Sci Rep}\ }\textbf {\bibinfo
  {volume} {7}},\ \bibinfo {pages} {1 -- 12} (\bibinfo {year}
  {2017})}\BibitemShut {NoStop}%
\bibitem [{\citenamefont {Digregorio}\ \emph {et~al.}(2018)\citenamefont
  {Digregorio}, \citenamefont {Levis}, \citenamefont {Suma}, \citenamefont
  {Cugliandolo}, \citenamefont {Gonnella},\ and\ \citenamefont
  {Pagonabarraga}}]{DigregorioPRL2018_fullphasediagramABP}%
  \BibitemOpen
  \bibfield  {author} {\bibinfo {author} {\bibfnamefont {P.}~\bibnamefont
  {Digregorio}}, \bibinfo {author} {\bibfnamefont {D.}~\bibnamefont {Levis}},
  \bibinfo {author} {\bibfnamefont {A.}~\bibnamefont {Suma}}, \bibinfo {author}
  {\bibfnamefont {L.~F.}\ \bibnamefont {Cugliandolo}}, \bibinfo {author}
  {\bibfnamefont {G.}~\bibnamefont {Gonnella}}, \ and\ \bibinfo {author}
  {\bibfnamefont {I.}~\bibnamefont {Pagonabarraga}},\ }\bibfield  {title}
  {\enquote {\bibinfo {title} {Full phase diagram of active brownian disks:
  From melting to motility-induced phase separation},}\ }\href {\doibase
  10.1103/PhysRevLett.121.098003} {\bibfield  {journal} {\bibinfo  {journal}
  {Phys. Rev. Lett.}\ }\textbf {\bibinfo {volume} {121}},\ \bibinfo {pages}
  {098003} (\bibinfo {year} {2018})}\BibitemShut {NoStop}%
\bibitem [{\citenamefont {Mazoyer}\ \emph {et~al.}(2009)\citenamefont
  {Mazoyer}, \citenamefont {Ebert}, \citenamefont {Maret},\ and\ \citenamefont
  {Keim}}]{mazoyer2009DynamicsParticlesCages}%
  \BibitemOpen
  \bibfield  {author} {\bibinfo {author} {\bibfnamefont {S.}~\bibnamefont
  {Mazoyer}}, \bibinfo {author} {\bibfnamefont {F.}~\bibnamefont {Ebert}},
  \bibinfo {author} {\bibfnamefont {G.}~\bibnamefont {Maret}}, \ and\ \bibinfo
  {author} {\bibfnamefont {P.}~\bibnamefont {Keim}},\ }\bibfield  {title}
  {\enquote {\bibinfo {title} {Dynamics of particles and cages in an
  experimental {{2D}} glass former},}\ }\href {\doibase
  10.1209/0295-5075/88/66004} {\bibfield  {journal} {\bibinfo  {journal} {EPL
  (Europhysics Letters)}\ }\textbf {\bibinfo {volume} {88}},\ \bibinfo {pages}
  {66004} (\bibinfo {year} {2009})}\BibitemShut {NoStop}%
\bibitem [{\citenamefont {Jacquin}\ and\ \citenamefont
  {Berthier}(2010)}]{Jacquin2010}%
  \BibitemOpen
  \bibfield  {author} {\bibinfo {author} {\bibfnamefont {H.}~\bibnamefont
  {Jacquin}}\ and\ \bibinfo {author} {\bibfnamefont {L.}~\bibnamefont
  {Berthier}},\ }\bibfield  {title} {\enquote {\bibinfo {title} {Anomalous
  structural evolution of soft particles: Equibrium liquid state theory},}\
  }\href {\doibase 10.1039/b926412d} {\bibfield  {journal} {\bibinfo  {journal}
  {Soft Matter}\ }\textbf {\bibinfo {volume} {6}},\ \bibinfo {pages} {2970}
  (\bibinfo {year} {2010})}\BibitemShut {NoStop}%
\bibitem [{\citenamefont {Berthier}, \citenamefont {Moreno},\ and\
  \citenamefont {Szamel}(2010)}]{berthier2010IncreasingDensityMelts}%
  \BibitemOpen
  \bibfield  {author} {\bibinfo {author} {\bibfnamefont {L.}~\bibnamefont
  {Berthier}}, \bibinfo {author} {\bibfnamefont {A.~J.}\ \bibnamefont
  {Moreno}}, \ and\ \bibinfo {author} {\bibfnamefont {G.}~\bibnamefont
  {Szamel}},\ }\bibfield  {title} {\enquote {\bibinfo {title} {Increasing the
  density melts ultrasoft colloidal glasses},}\ }\href {\doibase
  10.1103/PhysRevE.82.060501} {\bibfield  {journal} {\bibinfo  {journal}
  {Physical Review E}\ }\textbf {\bibinfo {volume} {82}},\ \bibinfo {pages}
  {060501} (\bibinfo {year} {2010})}\BibitemShut {NoStop}%
\bibitem [{\citenamefont {Ni}, \citenamefont {Stuart},\ and\ \citenamefont
  {Dijkstra}(2013)}]{Ni2013}%
  \BibitemOpen
  \bibfield  {author} {\bibinfo {author} {\bibfnamefont {R.}~\bibnamefont
  {Ni}}, \bibinfo {author} {\bibfnamefont {M.~A.~C.}\ \bibnamefont {Stuart}}, \
  and\ \bibinfo {author} {\bibfnamefont {M.}~\bibnamefont {Dijkstra}},\
  }\bibfield  {title} {\enquote {\bibinfo {title} {Pushing the glass transition
  towards random close packing using self-propelled hard spheres.}}\ }\href
  {\doibase 10.1038/ncomms3704} {\bibfield  {journal} {\bibinfo  {journal}
  {Nature communications}\ }\textbf {\bibinfo {volume} {4}},\ \bibinfo {pages}
  {2704} (\bibinfo {year} {2013})}\BibitemShut {NoStop}%
\bibitem [{\citenamefont {Philippe}\ \emph {et~al.}(2018)\citenamefont
  {Philippe}, \citenamefont {Truzzolillo}, \citenamefont {{Galvan-Myoshi}},
  \citenamefont {{Dieudonn\'e-George}}, \citenamefont {Trappe}, \citenamefont
  {Berthier},\ and\ \citenamefont
  {Cipelletti}}]{philippeGlassTransitionSoft2018}%
  \BibitemOpen
  \bibfield  {author} {\bibinfo {author} {\bibfnamefont {A.-M.}\ \bibnamefont
  {Philippe}}, \bibinfo {author} {\bibfnamefont {D.}~\bibnamefont
  {Truzzolillo}}, \bibinfo {author} {\bibfnamefont {J.}~\bibnamefont
  {{Galvan-Myoshi}}}, \bibinfo {author} {\bibfnamefont {P.}~\bibnamefont
  {{Dieudonn\'e-George}}}, \bibinfo {author} {\bibfnamefont {V.}~\bibnamefont
  {Trappe}}, \bibinfo {author} {\bibfnamefont {L.}~\bibnamefont {Berthier}}, \
  and\ \bibinfo {author} {\bibfnamefont {L.}~\bibnamefont {Cipelletti}},\
  }\bibfield  {title} {\enquote {\bibinfo {title} {Glass transition of soft
  colloids},}\ }\href {\doibase 10.1103/PhysRevE.97.040601} {\bibfield
  {journal} {\bibinfo  {journal} {Phys. Rev. E}\ }\textbf {\bibinfo {volume}
  {97}},\ \bibinfo {pages} {040601} (\bibinfo {year} {2018})}\BibitemShut
  {NoStop}%
\bibitem [{\citenamefont {Arceri}\ \emph {et~al.}(2020)\citenamefont {Arceri},
  \citenamefont {Landes}, \citenamefont {Berthier},\ and\ \citenamefont
  {Biroli}}]{arceriGlassesAgingStatistical2020}%
  \BibitemOpen
  \bibfield  {author} {\bibinfo {author} {\bibfnamefont {F.}~\bibnamefont
  {Arceri}}, \bibinfo {author} {\bibfnamefont {F.~P.}\ \bibnamefont {Landes}},
  \bibinfo {author} {\bibfnamefont {L.}~\bibnamefont {Berthier}}, \ and\
  \bibinfo {author} {\bibfnamefont {G.}~\bibnamefont {Biroli}},\ }\bibfield
  {title} {\enquote {\bibinfo {title} {Glasses and aging: {{A Statistical
  Mechanics Perspective}}},}\ }\href@noop {} {\bibfield  {journal} {\bibinfo
  {journal} {arXiv:2006.09725 [cond-mat]}\ } (\bibinfo {year} {2020})},\
  \Eprint {http://arxiv.org/abs/2006.09725} {arXiv:2006.09725 [cond-mat]}
  \BibitemShut {NoStop}%
\end{thebibliography}%

\end{document}